\documentclass[a4paper]{article}
\usepackage{color}
\usepackage{bm}
\usepackage{cite}
\usepackage{braket}
\usepackage{amsmath,amssymb,amsthm}
\usepackage[pdftex]{graphicx}

\def\abstract{\hfil{\large{\bf{Abstract}\vspace{-10pt}\\

}}}

\newcommand \beq{\begin{eqnarray}}
\newcommand \eeq{\end{eqnarray}}
\def\abf{a_{bf}^{\ }}

\def\obf{\omega_{_{bf}}^{\ }}
\def\gbf{g_{_{bf}}^{\ }}

\def\aF{a_{_{\rm FF}}}
\def\gF{g_{_{\rm FF}}}
\def\muF{\mu_{_{\rm F}}}
\def\mF{m_{_{\rm F}}}
\def\mb{m_{b}^{\ } }
\def\mf{m_{f}^{\ } }
\def\mR{m_{_{R}} }

\def\mub{\mu_{b}^{\ } }
\def\muf{\mu_{f}^{\ } }

\def\tr{\mathrm{tr}}
\def\omb{\omega_{b}^{\ } }
\def\ombf{\omega_{_{bf}}^{\ } }
\def\xib{\xi_{b}^{\ } }
\def\xif{\xi_{f}^{\ } }

\title{\textbf{Large $N$ Expansion for Strongly-coupled Boson-Fermion Mixtures \\ \vspace{40pt}}}
\author{Kenji Maeda\thanks{Email: kmaeda@nt.phys.s.u-tokyo.ac.jp}\vspace{5pt}\\
\textit{Department of Physics, University of Tokyo, Tokyo 113-0033, Japan}}

\date{revised in \today}

\begin{document}

\setlength{\baselineskip}{20pt}

\maketitle

\begin{abstract}
We study a many-body mixture of an equal number of bosons and two-component fermions  
with a strong contact attraction. 
In this system bosons and fermions can be paired into composite fermions. 
We construct a large $N$ extension 
where both bosons and fermions have the extra large $N$ degrees of freedom 
and the boson-fermion interaction is extended to a four-point contact interaction 
which is invariant under the ${\rm O}(N)$ group transformation, 
so that the composite fermions become 
singlet in terms of the ${\rm O}(N)$ group. 
It is shown that such ${\rm O}(N)$ singlet fields have controllable quantum fluctuations 
suppressed by $1/N$ factors and yield a systematic $1/N$-expansion in terms of composite fermions. 
We derive an effective action described by composite fermions up to the next-to-leading-order terms 
in the large $N$ expansion, and show that there can be the BCS superfluidity of composite fermions 
at sufficiently low temperatures. 

\end{abstract}
\newpage
\section{Introduction}

The study of boson-fermion mixtures has a long history originating 
from the analysis of dilute solutions of $^3$He atoms in superfluid $^4$He \cite{BBP67}. 
For a weakly-coupled boson-fermion mixture, it is known that 
the density fluctuation of the bosonic background induces an attraction between the fermions, 
which enhances the transition temperature to the BCS superfluidity or leads to fermionic superfluidity 
even without a bare attractive potential between fermions \cite{BHS00,HPSV00}. 
On the other hand, in the strong coupling regime, 
it is possible to form bound states between bosons and fermions, 
called composite fermions (CFs) or simply dimers (tightly-bound molecules) \cite{SSSYD05}. 
Therefore, phase structures 
in the strong coupling regime 
may differ from those in the weak coupling regime, 
and it is expected that there occurs the superfluidity of CFs 
at low temperatures \cite{KBEK04}, which greatly motivates us to model 
superfluid hadronic matters in dense QCD in terms of boson-fermion mixtures 
where small size diquarks correspond to the bosons, unpaired quarks to the fermions, 
and the extended nucleons are regarded as the CFs \cite{BHTY08, MBH09, HM09}. 

Recent developments in atomic experiments 
have made it possible to realize boson-fermion mixed gases in the laboratory. 
Atomic interaction between different species can be tuned 
with the use of Feshbach resonance techniques \cite{FES58,KGJ06,BDZ08}. 
Recently, the formation of heteronuclear Feshbach molecules has been 
observed in a boson-fermion mixture of $^{87}$Rb and $^{40}$K atomic vapors 
in a 3D optical lattice \cite{OSP06} and in an optical dipole trap \cite{ZIR08}. 

From a theoretical point of view, there are several non-perturbative studies 
on nonrelativistic atomic gases. In particular, the large $N$ method provides 
a systematic expansion with the corresponding diagrammatic representations, 
and the applicability of its results to the physical cases at $N=1$ 
can in principle be tested by systematic estimates of higher-order contributions. 
The transition temperature of the dilute interacting Bose gas has been calculated 
with the use of $1/N$-expansion \cite{BBZ00,AMT01}. 
Also, the $1/N$-expansion for the nonrelativistic Fermi gases has been developed in 
Refs.\cite{NS07,VSR07,AB08}. 
A review of large $N$ expansions in ${\rm O}(N)$ and ${\rm U}(N)$ quantum field theories, 
which deals with non-perturbative aspects of critical phenomena,  
may be found in Ref.\cite{MZJ2003}. 
However, detailed studies of the strongly-coupled boson-fermion mixtures in the large $N$ method 
are still missing. 

In this paper we present an extensive study of a large $N$ extension for a model of 
strongly-coupled boson-fermion mixtures originally proposed in Ref.\cite{MBH09}. 
We establish the $1/N$-expansion in a theory of CFs 
which is equivalent to the original boson-fermion mixed system. 
We also derive an effective action of CFs 
up to the next-to-leading-order terms in the large $N$ expansion, 
and show that there can be the BCS superfluidity of CFs 
at low temperatures. 

Our paper is organized as follows. 
In Sect.\:2, we construct a large $N$ extension of strongly-coupled boson-fermion mixtures 
at finite temperature and density based on the imaginary-time formalism. 
In Sect.\:3, we rewrite the boson-fermion partition function 
in terms of CFs with the use of an auxiliary-field method. 
We derive an action functional of CFs 
and find a systematic expansion, $1/N$-expansion, 
which is equivalent to a loop expansion with respect to the CF fields.
In Sect.\:4, the $1/N$-expansion is employed to calculate 
the leading-order (LO) and the next-to-leading-order (NLO) terms in our CF action. 
We also derive a low-energy effective theory of CFs, 
and find that it reduces to a two-component free Fermi gas in the LO analysis, 
and to a weakly-interacting two-component Fermi gas up to the NLO study, 
which yields the superfluidity of CFs at sufficiently low temperatures. 
Finally, in Sect.\:5 we discuss 
the application of boson-fermiuon mixtures to dense QCD. 
In Appendix A, we give explicit forms of Fourier transformations especially 
for the proper vertex functions of CF fields. 
Appendix B provides details on the derivative expansion of 
the inverse propagator of CFs. 

\section{Formulation of large $N$ boson-fermion mixtures}

In our model, we treat bosons and two-component fermions 
using a nonrelativistic gas model of the boson-fermion mixture
where bosons and fermions interact through a four-point contact interaction. 
We start from a Hamiltonian density of our boson-fermion mixture in three spatial dimensions, 
\beq
\mathcal{H}
&=& 
\sum_{i=1}^N 
\phi_i ^{\dagger}(x)
\biggl(-\frac{\nabla ^2}{2\mb}-\mub \biggl)
\phi_i (x)  
\:+\: 
\sum_{i=1}^N \sum_{\sigma=\uparrow,\downarrow} 
\psi_{\sigma i}^{\dagger} (x)
\biggl(-\frac{\nabla ^2}{2\mf}-\muf \biggl)
 \psi_{\sigma i}(x) 
 \nonumber\\
 &&+~
\frac{\gbf}{N}\sum_{i,j=1}^N \sum_{\sigma=\uparrow,\downarrow}
\phi_i^{\dagger}(x)\psi_{\sigma i}^{\dagger}(x) \phi_j(x) \psi_{\sigma j}(x) \: ,
\label{def:H}
\eeq
where $\phi_i$ is the bosonic and $\psi_{\sigma i}$ is the fermionic field. 
We label the two internal states of the fermions by pseudospin indices $\sigma=\uparrow, \downarrow$ 
and extra large $N$ indices of bosons and fermions by $i,j=1,2,\dots , N$. 
We assume that two different pseudospin states have 
the same mass and chemical potential (number density) 
and that the boson-fermion interaction is independent of pseudospin states. 
Setting $N=1$ yields the same Hamiltonian density as in our previous work  \cite{MBH09}, 
though here we neglect interactions between same species by assuming that 
the boson-fermion interaction is much stronger than the others. 
To make our analysis simple, we focus on an equally populated mixture of bosons and fermions, 
which means for each $i$ we have $n$ bosons and $n$ fermions 
with an equal population in their number densities: $n_{bi}=n_{\uparrow i} + n_{\downarrow i}=n$ and 
$n_{\uparrow i}=n_{\downarrow i}=n/2$. 
Also, we introduce total boson (fermion) number density as $n_{\rm tot}=Nn$. 

The bare boson-fermion coupling $\gbf$ is related to 
the s-wave scattering length in the vacuum $\abf$ by the following relation \cite{PS2008}, 
\beq
\frac{\mR}{2\pi \abf} 
&=&\frac{1}{\gbf}+
 \int_{|{\bf k}| \le \Lambda}\frac{d{\bf k}}{(2\pi)^3} 
 \frac{1}{\varepsilon_{b}({\bf k})+ \varepsilon_{f}({\bf k})}\: ,  
\label{swave}
\eeq 
where $\varepsilon_{b}({\bf k})= {\bf k}^2/2\mb$ and $\varepsilon_{f}({\bf k})= {\bf k}^2/2\mf$ are the kinetic energies 
of the single boson and fermion, respectively, $\mR=\mb\mf/(\mb+\mf)$ is the boson-fermion reduced mass, 
and $\Lambda$ is a high-momentum cutoff of our model which sets a minimum atomic scale $r_0=(2\Lambda/\pi)^{-1}$. 
For a simple notation, we will omit the constrain to the momentum-integral; $|{\bold k}| \le \Lambda$. 

\begin{figure}[t]
\begin{center}
\includegraphics[width=11cm]{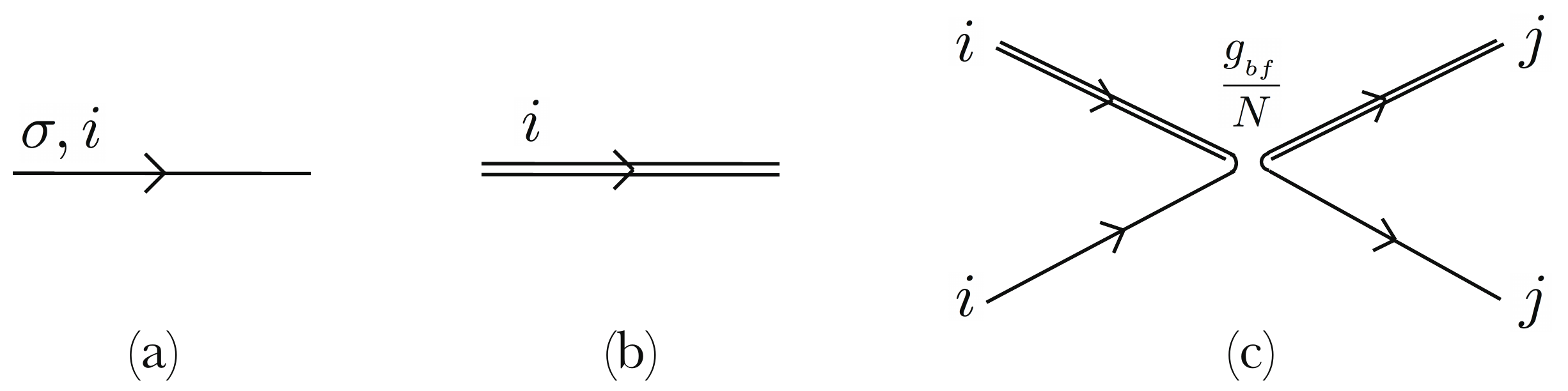}
\end{center}
\vspace{-0.5cm}
\caption{Feynman diagrams for the boson-fermion mixture described by the action Eq.(\ref{def:S}). 
The indices for large $N$ degrees of freedom are explicitly shown by $i$ or $j$. 
(a) The single line denotes one of propagators for two-component fermions labeled by $\sigma$ and $i$,  
(b) the double line corresponds to the propagator for bosons, 
and (c) the interaction vertex between bosons and fermions is represented by 
the empty space associated with a suppression factor $\gbf/N$ 
in terms of the $1/N$-expansion. 
}
\label{fig1}
\end{figure}

The partition function at finite temperature becomes 
\beq
Z&=&\int\biggl(\: \prod_{i=1}^N {\cal D}\phi_i ^{\ast}{\cal D}\phi_i  \biggl)
\biggl(\: \prod_{i=1}^N \prod_{\sigma=\uparrow,\downarrow} 
{\cal D}\bar{\psi}_{\sigma i} {\cal D}\psi_{\sigma i} \biggl)
\exp \biggl\{ 
-S\bigl[\phi_i^{\ast}, \phi_i,\bar{\psi}_{\sigma i}, \psi_{\sigma i}\bigl] 
\biggl\} \: ,
\label{def:Z}
\eeq
expressed by an imaginary-time functional integral over bosonic fields $\phi_i,\phi_i  ^{\ast}$ 
and fermionic Grassmann fields $\psi_{\sigma i},\bar{\psi}_{\sigma i}$ 
with the corresponding action functional of our boson-fermion mixture: 
\beq
S\bigl[\phi_i^{\ast}, \phi_i,\bar{\psi}_{\sigma i}, \psi_{\sigma i}\bigl] 
&=&
 \int\! dx\: \sum_{i=1}^N
\phi_i ^{\ast}(x)
\biggl(\frac{\partial }{\partial \tau}-\frac{\nabla ^2}{2\mb}-\mub \biggl)
\phi_i (x) 
\nonumber\\ 
&&+~
\int\! dx\: \sum_{i=1}^N \sum_{\sigma=\uparrow,\downarrow} 
\bar{\psi}_{\sigma i} (x)
\biggl(\frac{\partial }{\partial \tau}-\frac{\nabla ^2}{2\mf}-\muf \biggl)
 \psi_{\sigma i}(x)
\nonumber\\ 
&&+~
\int\! dx \:\frac{\gbf}{N}\sum_{i,j=1}^N \sum_{\sigma=\uparrow,\downarrow}
\phi_i^{\ast}(x)\bar{\psi}_{\sigma i}(x)\phi_j(x) \psi_{\sigma j}(x) \: .
\label{def:S}
\eeq
Here  for simple descriptions we have used notations: $x=(\textbf{x},\tau)$, 
$\int\! dx= \int_0 ^{\beta}d\tau\int\! d\textbf{x}$, and $\hbar=k_B=1$. 
Figure \ref{fig1} shows the corresponding Feynman diagrams for the boson-fermion mixture, 
especially focused on the large $N$ degrees of freedom.

\section{From boson-fermion mixtures to composite fermions}

\begin{figure}[t]
\begin{center}
\includegraphics[width=15cm]{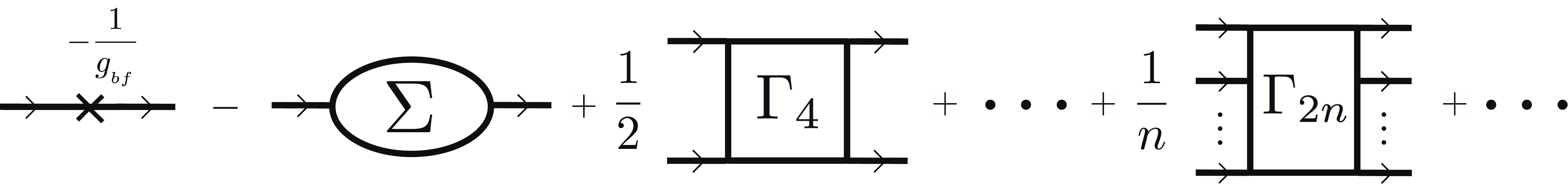}
\end{center}
\vspace{-0.5cm}
\caption{Graphical representation of the CF action 
${\cal S}\bigl[\bar{\cal F}_{\sigma},{\cal F}_{\sigma}\bigl]$ in Eq.(\ref{1/N_full_action}). 
The thick lines represent the external lines of 
CF fields $\bar{\cal F}_{\sigma}$ and ${\cal F}_{\sigma}$. 
Definitions of $\Sigma$, $\Gamma_4$ and $\Gamma_{2n}$ will be given 
later in Eq.(\ref{F_self}), (\ref{4v_full}) and (\ref{2nv_full}), respectively. 
}
\label{fig_F_Fey}
\end{figure}

In strongly-coupled boson-fermion mixtures characterized by 
the positive and small scattering length 
($0< n_{\rm tot}^{1/3}\abf \ll 1$), 
we expect that bosons and fermions form bound dimers 
or {\it composite fermions} (CFs) \cite{SSSYD05,KBEK04} and that 
low-energy phenomena can be described by an effective theory of these CFs.  
For this purpose, 
we introduce fermionic auxiliary fields $F^{\prime}_{\sigma}(x)$ and $\bar{F}^{\prime} _{\sigma}(x)$ 
by inserting the following identity into the partition function: 
\beq
1&=& c\int \biggl(\: \prod_{\sigma=\uparrow,\downarrow}
{\cal D}\bar{F}^{\prime} _{\sigma} {\cal D}F_\sigma ^{\prime}\biggl)\:
\mathrm{exp}\biggl\{\frac{N}{\gbf}\int\! dx \sum_{\sigma=\uparrow,\downarrow}
\bar{F}^{\prime}_\sigma(x)F_\sigma ^{\prime}(x) \biggl\} \:,
\label{FHS}
\eeq
where $c$ is a normalization constant \cite{RHET02}. We also define shifted fields 
$F_\sigma(x)$ and $\bar{F}_{\sigma}(x)$ as
\beq
F^{\prime}_\sigma (x)&=&
\frac{\gbf}{N}\sum_{i=1}^N \phi_i(x)\psi_{\sigma i} (x)\:+\:  F_\sigma (x) \; ,\\
\bar{F}^{\prime}_{\sigma}(x)&=&
\frac{\gbf}{N}\sum_{i=1}^N\phi^{\ast}_i(x)\bar{\psi}_{\sigma i}(x)\:+\:\bar{F}_\sigma(x) \: . 
\label{shiftedF}
\eeq
Note that the shifted fields 
$F_\sigma(x)$ and $\bar{F}_{\sigma}(x)$ can be considered as 
fluctuations of $F^{\prime}_\sigma$ and $\bar{F}^{\prime}_{\sigma}$ 
around O$(N)$ singlet fields $\sum_{i=1}^N \phi_i(x)\psi_{\sigma i} (x)/N$ 
and $\sum_{i=1}^N\phi^{\ast}_i(x)\bar{\psi}_{\sigma i}(x)/N$ respectively, both 
of which are arithmetic averages of many fields in terms of the large $N$. 
As we will see below we can in principle control these fluctuations 
by changing $N$ itself. 
The partition function Eq.(\ref{def:Z}) then becomes 
\beq
Z&=&\!\! c\!\int\! 
{\cal D}[\bar{F}_{\sigma}^{\prime},F^{\prime}_{\sigma}]
{\cal D}[\phi_i^{\ast},\phi_i]
{\cal D}[\bar{\psi}_{\sigma i},\psi_{\sigma i}]
\nonumber\\
&& \mathrm{exp} \biggl\{ \int\! dxdy \biggl[~
\sum_{i=1}^N \phi_i^{\ast}(x)D^{-1}(x,y)\phi_i(y) + \sum_{i=1}^N\sum_{\sigma=\uparrow,\downarrow}
 \bar{\psi}_{\sigma i}(x)S^{-1}(x,y)\psi_{\sigma i}(y)
\biggl] 
\nonumber \\
& &
- \int\! dx\biggl[~ \sum_{i,j=1}^N\sum_{\sigma=\uparrow,\downarrow}\frac{\gbf}{N} \phi^{\ast}_i(x) \bar{\psi}_{\sigma i}(x)
\phi_j(x) \psi_{\sigma j}(x)
-\frac{N}{\gbf}\sum_{\sigma=\uparrow,\downarrow}\bar{F}^{\prime}_\sigma(x) F_\sigma ^{\prime}(x)\biggl] \biggl\} \nonumber \\ 
 &=&\!\! c\!\int\! 
{\cal D}[\bar{F}_{\sigma},F_{\sigma}]
{\cal D}[\phi_i^{\ast},\phi_i]
{\cal D}[\bar{\psi}_{\sigma i},\psi_{\sigma i}]
\nonumber\\ &&
 \mathrm{exp} \biggl\{ \int\! dxdy \biggl[~
 \sum_{i=1}^N \phi_i^{\ast}(x)D^{-1}(x,y)\phi_i(y) + \sum_{i=1}^N\sum_{\sigma=\uparrow,\downarrow}
 \bar{\psi}_{\sigma i}(x)S^{-1}(x,y)\psi_{\sigma i}(y)
\biggl] \nonumber \\
& &
+\!\int\! dx \sum_{\sigma=\uparrow,\downarrow} \:\biggl[~  \frac{N}{\gbf}\bar{F}_{\sigma}(x)F_{\sigma} (x)
+\sum_{i=1}^N\bar{F}_\sigma(x)\phi_i(x)\psi_{\sigma i}(x)
+\sum_{i=1}^N\phi_i^{\ast}(x)\bar{\psi}_{\sigma i}(x)F_{\sigma}(x) \biggl] \biggl\} \: ,
\label{Z_FHS}
\eeq
where $D^{-1}(x,y)$ and $S^{-1}(x,y)$ denote inverse Green's functions of bosons and fermions, respectively: 
\beq
D^{-1}(x,y)&=&\biggl(-\partial _{\tau}+\frac{\nabla ^2}{2m_{b}}+\mu_{b} \biggl)\delta(x-y)\: , \\
S^{-1}(x,y)&=&\biggl(-\partial _{\tau}+\frac{\nabla ^2}{2m_{f}}+\mu_{f} \biggl)\delta(x-y) \: .
\label{def:IGF}
\eeq
We also used a simple notation for functional integral measures: 
\beq
{\cal D}[\bar{F}_{\sigma},F_{\sigma}]
{\cal D}[\phi_i^{\ast},\phi_i]
{\cal D}[\bar{\psi}_{\sigma i},\psi_{\sigma i}]
\!\!\!&=&\!\!\!
\biggl(\: \prod_{\sigma=\uparrow,\downarrow}
{\cal D}\bar{F}_{\sigma} {\cal D}F_{\sigma}\biggl)
\biggl(\: \prod_{i=1}^N {\cal D}\phi_i ^{\ast}{\cal D}\phi_i  \biggl)
\biggl(\: \prod_{i=1}^N \prod_{\sigma=\uparrow,\downarrow}
{\cal D}\bar{\psi}_{\sigma i}{\cal D}\psi_{\sigma i} \biggl) \:. 
\eeq
Note that as in a usual Hubbard-Stratonovich transformation 
the introduction of auxiliary fields reduces the original bosonic and fermionic fields 
into bilinear forms which are diagonalized in terms of the large $N$ indices, and 
they can be integrated out immediately
. 
We first perform the fermionic functional integral in Eq.(\ref{Z_FHS}), 
and the relevant part of the integration yields 
for each pair of indices $\sigma=\uparrow,\downarrow$ and 
$i=1,\dots ,N$ ({\it i.e.}, summation over $\sigma$ and $i$ is not assumed here), 
\beq
\lefteqn{ \int\!{\cal D}\bar{\psi}_{\sigma i}{\cal D}\psi_{\sigma i}
\mathrm{exp} \biggl\{ \int\! dxdy\left[
\bar{\psi}_{\sigma i}(x)S^{-1}(x,y)\psi_{\sigma i}(y)
+\bar{F}_\sigma(x)\phi_i(x)\delta(x-y)\psi_{\sigma i}(y)
+\phi_i^{\ast}(x)\delta(x-y)\bar{\psi}_{\sigma i}(y)F_{\sigma (y)} \right] \biggl\} }
 \nonumber\\
 &=&\! \int\!{\cal D}\bar{\psi}_{\sigma i}{\cal D}\psi_{\sigma i}
 \mathrm{exp}\biggl\{ \int\! dxdy 
 \bigl[ \bar{\psi}_{\sigma i}(x)\!+\!\int\! dz \bar{F}_{\sigma}(z)\phi_i(z)S(z,x)\bigl]S^{-1}(x,y)
 \bigl[\psi_{\sigma i}(y)\!+\!\int\! dw S(y,w)\phi_i^{\ast}(w)F_\sigma (w) \bigl] \nonumber \\
 & &\qquad \qquad  \qquad  \qquad 
-\int\! dxdydzdw 
\bar{F}_{\sigma}(z)\phi_i(z)S(z,x)S^{-1}(x,y)S(y,w)\phi_i^{\ast}(w)F_{\sigma} (w) \biggl\} 
 \nonumber \\
 &=&\! \int\!{\cal D}\bar{\psi}_{\sigma i}^{\prime}{\cal D}\psi_{\sigma i}^{\prime}
 \mathrm{exp}
 \biggl\{\int\! dxdy \: \bar{\psi}^{\prime} _{\sigma i}(x)S^{-1}(x,y)\psi^{\prime} _{\sigma i}(y) \biggl\}\: 
 \mathrm{exp}\biggl\{-\int\! dzdw 
 \bar{F}_{\sigma}(z)\phi_i(z)S(z,w)\phi_i^{\ast}(w)F_\sigma (w) \biggl\}
 \nonumber \\
 &=&\!\mathrm{exp}\biggl\{\ln\det \bigl[-S^{-1}(x,y)\bigl] -\int\! dxdy \:\phi_i^{\ast}(x)\bar{F}_{\sigma}(y)S(y,x)
 F_\sigma (x)\phi_i(y) \biggl\}
 \nonumber \\
 &=&\!\mathrm{exp}\biggl\{\mathrm{tr}\ln \bigl[-S^{-1}(x,y)\bigl]
 - \int\! dxdy \:\phi_i^{\ast}(x) A_{\sigma}(x,y)\phi_i(y)\biggl\}\:,
\label{f_int}
\eeq
where ``$\tr$" and ``$\det$" are taken only over coordinate indices, and 
$A_{\sigma}(x,y)$ is defined by $A_{\sigma}(x,y)= \bar{F}_{\sigma}(y)S(y,x)F_\sigma (x)$, 
whose ordering is important due to the anti-commuting nature of Grassmann fields $\bar{F}_{\sigma}$ and $F_\sigma$. 
We have also used a matrix formula: $\ln\det M=\mathrm{tr}\ln M$. 
Thus, Eq.(\ref{Z_FHS}) reduces to
\beq
Z&=& c \bigl[Z_0^f(\muf)\bigl]^N \!\int\!{\cal D}\bigl[\bar{F}_{\sigma},F_{\sigma}\bigl]
{\cal D}\bigl[\phi_i^{\ast},\phi_i\bigl]
\label{Z-f}\\
 &&
\mathrm{exp}\biggl\{\int\! dxdy\:
\sum_{i=1}^N\phi_i^{\ast}(x)\biggl[ D^{-1}(x,y)-\sum_{\sigma=\uparrow,\downarrow}A_{\sigma}(x,y)\biggl]\phi_i(y)
 +\frac{N}{\gbf}\int\! dx \sum_{\sigma=\uparrow,\downarrow}\bar{F}_{\sigma}(x)F_\sigma (x) \biggl\} \:, \nonumber
\eeq
with a partition function for a two-component free Fermi gas 
$Z_0^f(\muf)= \exp\bigl\{2\mathrm{tr}\ln [-S^{-1}(x,y)]\bigl\}$. 
We proceed to perform bosonic functional integral, 
and the relevant part of the integration yields for each index $i=1,\dots ,N$, 
\beq
\lefteqn{ 
\int\!{\cal D}\phi_i^{\ast}{\cal D}\phi_i\:
\mathrm{exp}\biggl\{\int\! dxdy \:
\phi_i^{\ast}(x)\biggl[ D^{-1}(x,y)-\sum_{\sigma=\uparrow,\downarrow}A_{\sigma}(x,y)\biggl]\phi_i(y)\biggl\}
} \nonumber\\
&=&\biggl\{ \det \biggl[-D^{-1}(x,y)+\sum_{\sigma=\uparrow,\downarrow}A_{\sigma}(x,y)\biggl]\biggl\}^{-1} 
\nonumber\\
&=&\biggl\{\det \int\! dz\bigl[-D^{-1}(x,z)\bigl]
\biggl[\delta (z-y)-\sum_{\sigma=\uparrow,\downarrow} \int\! dwD(z,w)A_{\sigma}(w,y)\biggl]\:\biggl\}^{-1} 
\nonumber \\
 &=&\mathrm{exp}\biggl\{-\mathrm{tr}\ln \bigl[-D^{-1}(x,y)\bigl]
 -\mathrm{tr}\ln \biggl[\delta (x-y)-\sum_{\sigma=\uparrow,\downarrow}\int\! dwD(x,w)A_{\sigma}(w,y)\biggl]\biggl\}\:.
\label{b_int}
\eeq
Applying Eq.(\ref{b_int}) to Eq.(\ref{Z-f}), we obtain a partition function 
which is described only by CF fields: 
\beq
Z &=& c \bigl[ Z_0^b(\mub) Z_0^f(\muf)\bigl]^N \\
&&\!\!\! \times \int\!{\cal D}[\bar{F}_{\sigma},F_{\sigma}]
\mathrm{exp}\biggl\{\frac{N}{\gbf}\int\! dx \sum_{\sigma=\uparrow,\downarrow}\bar{F}_{\sigma}(x)F_\sigma (x) 
-N \mathrm{tr}\ln \biggl[\delta (x-y)-\sum_{\sigma=\uparrow,\downarrow}\int\! dwD(x,w)A_{\sigma}(w,y)\biggl] \biggl\} \:,
\nonumber
\label{Z_F}
\eeq
with a partition function for the ideal Bose gas 
$Z_0^b(\mub)= \mathrm{exp} \bigl\{-\mathrm{tr}\ln \bigl[-D^{-1}(x,y)\bigl]\bigl\}$. 
The corresponding action becomes  
\beq
S[\bar{F}_{\sigma},F_{\sigma}]\!\!\!&=&\!\!\!
-\frac{N}{\gbf}\int\! dx \sum_{\sigma=\uparrow,\downarrow}\bar{F}_{\sigma}(x)F_\sigma (x)
+N\mathrm{tr}\ln \biggl[\delta (x-y)-\sum_{\sigma=\uparrow,\downarrow}\int\! dwD(x,w)A_{\sigma}(w,y)\biggl] \:.
\label{S_F}
\eeq
Since $N$ becomes an overall factor in the action 
and plays the same role as $\hbar$ in a usual loop expansion, 
our $1/N$-expansion is equivalent to the loop expansion based on 
the CF action Eq.(\ref{S_F})\footnote{In general when we have 
an overall factor $1/r$ in our action, 
propagators should be proportional to $r$, while any kind of vertices to $1/r$. 
Then, any graph composed of $P$ propagators and $V$ vertices 
is proportional to $r^{P-V}$. On the other hand, 
such a graph has $L=P-(V-1)$ loops, which yields a relation: $r^{P-V}=r^{L-1}$. 
Therefore, the series expansion in terms of $r$ is equivalent to 
the loop expansion in diagrammatic expressions\cite{NO1988}. 
Setting $1/r=N$ yields our $1/N$-expansion.}. 
Let us normalize the CF fields as 
\beq
{\cal F}_{\sigma}(x)&=&\sqrt{N} F_{\sigma}(x) \:, \\
\bar{\cal F}_{\sigma}(x)&=&\sqrt{N}\bar{F}_{\sigma}(x)\: , \\
{\cal A}_{\sigma}(x,y)&=&\bar{\cal F}_{\sigma}(y)S(y,x){\cal F}_\sigma (x)
~=~ NA_{\sigma}(x,y) \: ,
\eeq
which give an explicit form of the $1/N$-expansion, 
with the use of a formula in the logarithm: $\ln (1-M)=-\sum_{k=1}^{\infty}M^k/k$, 
\beq
S[\bar{F}_{\sigma},F_{\sigma}]
\!\!\!\!&=&\!\!\!\!
-\frac{1}{\gbf}\int\! dx \sum_{\sigma=\uparrow,\downarrow}\bar{\cal F}_{\sigma}(x){\cal F}_{\sigma}(x)
+N\mathrm{tr}\ln \biggl[\delta (x-y)
-\frac1{N}\int\! dw \sum_{\sigma=\uparrow,\downarrow} D(x,w){\cal A}_{\sigma}(w,y)\biggl] 
\nonumber\\  
\!\!\!\!&=&\!\!\!\!
-\frac{1}{\gbf }\int\! dx \sum_{\sigma=\uparrow,\downarrow}\bar{\cal F}_{\sigma}(x){\cal F}_{\sigma}(x)
-\sum_{k=1}^{\infty}\biggl(\frac{1}{ N}\biggl)^{k-1} \frac{1}{k} \mathrm{tr}
\biggl\{\int\! dw \sum_{\sigma=\uparrow,\downarrow} D(x,w)\bar{\cal F}_{\sigma}(y)S(y,w){\cal F}_\sigma (w)\biggl\}^k
\nonumber\\
\!\!\!\!&=:&\!\!
{\cal S}\bigl[\bar{\cal F}_{\sigma},{\cal F}_{\sigma}\bigl] \: .
\label{1/N_full_action}
\eeq
Here one can see that there is no internal degree of freedom 
associated with the large $N$ extension and that 
$1/N$ only appears as a suppression factor of 
each higher-dimensional interaction between CFs. 
Then, we reach the following representation of the partition, 
\beq
Z&=& c' \bigl[ Z_0^b(\mub) Z_0^f(\muf)\bigl]^N
 \int\!{\cal D}[\bar{\cal F}_{\sigma},{\cal F}_{\sigma}]
\mathrm{exp}\bigl\{-{\cal S}\bigl[\bar{\cal F}_{\sigma},{\cal F}_{\sigma}\bigl]\bigl\}~, 
\eeq
 with a normalization constant $c'$. 
Figure \ref{fig_F_Fey} shows a formal expression of 
the CF action ${\cal S}\bigl[\bar{\cal F}_{\sigma},{\cal F}_{\sigma}\bigl]$ 
of Eq.(\ref{1/N_full_action}) in terms of Feynman graphs.  
We will give precise definitions of $\Sigma$ (``self-energy"), 
$\Gamma_4$ (4-point vertex function) and 
$\Gamma_{2n}$ ($2n$-point vertex function) 
later in Eq.(\ref{F_self}), (\ref{4v_full}) and (\ref{2nv_full}), respectively. 
It will be shown that the first two graphs in Fig.\ref{fig_F_Fey} yields 
an inverse propagator of CFs which behaves as a free Fermi particle 
within our approximation, and 
the rest of graphs can be considered as interaction vertices of CFs. 

\section{$1/N$-expansion of strongly-coupled boson-fermion mixtures}

In the following, 
we will perform the $1/N$-expansion based on Eq.(\ref{1/N_full_action}) 
up to the next-to-leading-order terms 
and derive a low-energy effective theory of CFs. 
We will show that under an assumption discussed below 
an effective interaction between CFs are weakly attractive, 
and that the BCS-superfluidity of CFs (CF-BCS) is realized at sufficiently low temperatures. 

\subsection{The leading-order terms}
\begin{figure}[t]
\begin{center}
\includegraphics[width=10cm]{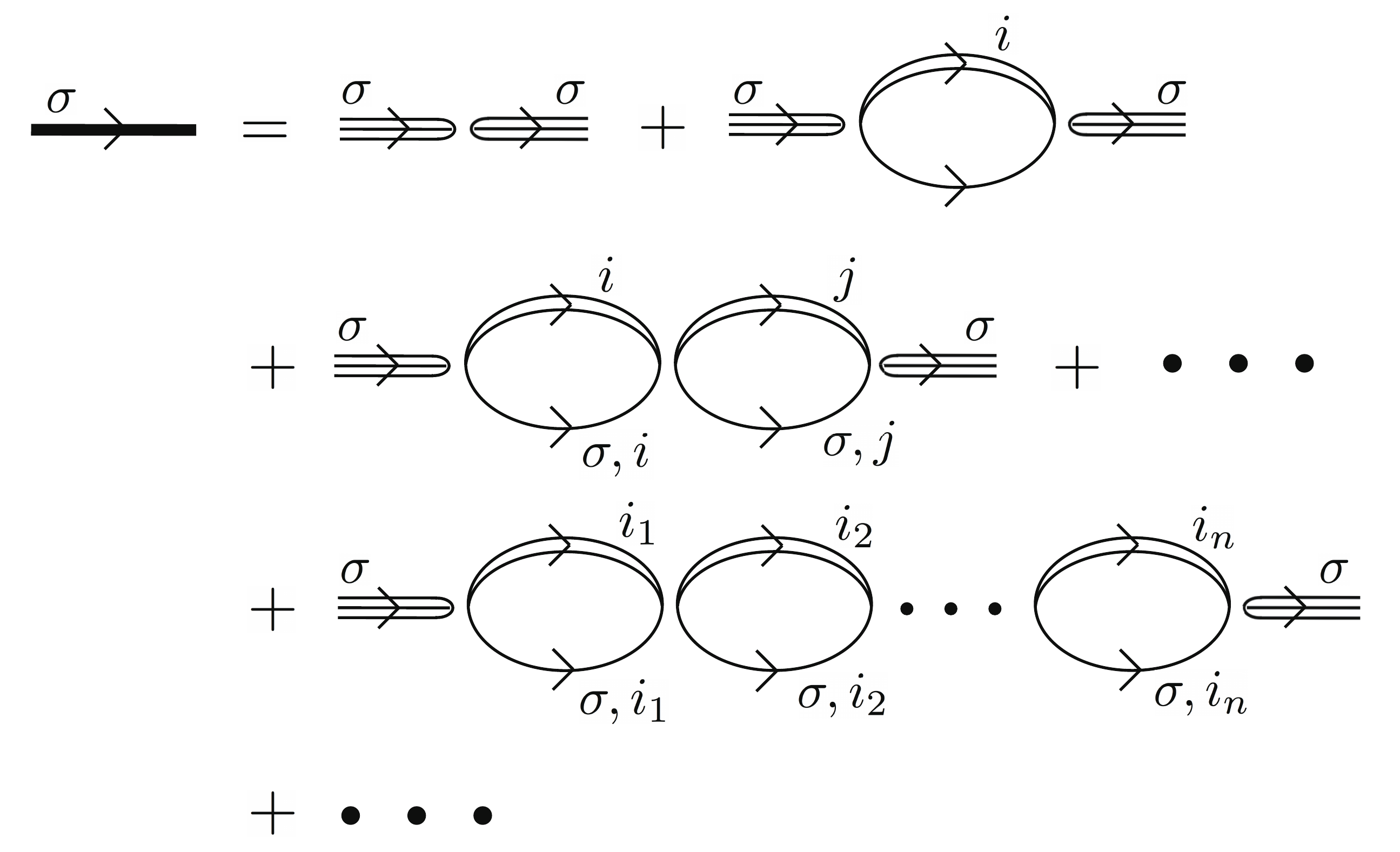}
\end{center}
\vspace{-0.5cm}
\caption{Graphical representation of the 
propagator for ``bare" CF fields 
$F_{\sigma}$ (not for ${\cal F}_{\sigma}$). 
In the right hand side, the external triple line denotes a pair of single boson and fermion, 
and only appears as a shorthand notation for a half of the vertex in Fig.\:\ref{fig1}(c). 
As for the propagator of ${\cal F}_{\sigma}(:= \sqrt{N}F_{\sigma})$ fields, 
we need to multiply both sides by the normalization factor $N$.
}
\label{fig2}
\end{figure}

The leading-order (LO) terms in the $1/N$-expansion, 
{\it i.e.}, $O(1)$ terms in Eq.(\ref{1/N_full_action}), become
 \beq
 {\cal S}_{\rm LO}\bigl[\bar{\cal F}_{\sigma},{\cal F}_{\sigma}\bigl]
\!\!&=&\!\!-\frac{1}{\gbf }\int\! dx \sum_{\sigma=\uparrow,\downarrow}\bar{\cal F}_{\sigma}(x){\cal F}_{\sigma}(x)
-\int\! dxdw \sum_{\sigma=\uparrow,\downarrow}\bar{\cal F}_{\sigma}(x)D(x,w)S(x,w){\cal F}_{\sigma} (w) \:,
\label{LO_x}
\eeq
which we can rewrite with the use of Fourier transforms (see Appendix \ref{FT}) as, 
\beq
{\cal S}_{\rm LO}\bigl[\bar{\cal F}_{\sigma},{\cal F}_{\sigma}\bigl]
&=&-\frac{1}{\gbf }\int\! dp \sum_{\sigma=\uparrow,\downarrow}\bar{\cal F}_{\sigma}(p){\cal F}_{\sigma}(p)
-\int\! dp\sum_{\sigma=\uparrow,\downarrow}\bar{\cal F}_{\sigma}(p)
\biggl[\:\int\! dq D(q)S(p-q)\biggl]{\cal F}_{\sigma}(p)
\nonumber \\
&=&
- \int\! dp\sum_{\sigma=\uparrow,\downarrow}\bar{\cal F}_{\sigma}(p)G^{-1}(p) {\cal F}_{\sigma}(p)~.
\label{LO_p}
\eeq
Here we have used notations: $p=(\textbf{p},i\omega)$, 
$\int\! dp= T\sum_{\omega}\omega \int\! d\textbf{p}/(2\pi)^3$ 
with the Matsubara frequency $\omega$ and spatial momentum vector ${\bold p}$. 
Also, we have introduced an inverse propagator of CF fields ${\cal F}_{\sigma}$ as 
\beq
G^{-1}(p) &=&\frac{1}{\gbf}  \:+\: \Sigma(p) \:,
\label{F_prop}
\eeq
with a CF ``self-energy", or single ``bubble" of bosons and fermions, 
$\Sigma(p)$ given by 
\beq
\Sigma(p)&=& \int\! dq D(q)S(p-q) \: .
\label{F_self}
\eeq
Equations (\ref{F_prop}) and (\ref{F_self}) show that 
the propagator is represented by 
an infinite geometric series of the original boson-fermion bubbles, as shown in Fig.\:\ref{fig2}. 
In the right hand side of Fig.\:\ref{fig2}, we can see that 
the $n$-th graph has large $N$ power-counting factors (i) $N^{n-1}$ from $n-1$ internal loops, 
(ii) $(1/N)^{n}$ from $n$ vertices, and (iii) $(\sqrt{N})^{2}$ from the normalization 
(${\cal F}_{\sigma}= \sqrt{N}F_{\sigma}$) for any $n\in\mathbb{N}$, 
which give an $O(1/N^{0})$ term in total, 
{\it i.e.}, the leading-order contribution in the $1/N$-expansion 
as shown in Eqs.(\ref{LO_p})-(\ref{F_self}). 

Let us expand the inverse propagator $G^{-1}$ 
in order to derive a low-energy effective theory of CFs. 
The summation over the bosonic Matsubara frequency 
in the self-energy Eq.(\ref{F_self}) is performed as 
\beq
\Sigma(p)&=&\int\! dq D(q)S(p-q) \nonumber\\
&=&T\sum_{\omb}\int\!\frac{d {\bf q}}{(2\pi)^3}
\frac{1}{i\omb -\xib ({\bf q})}\frac{1}{i(\omega-\omb) -\xif ( {\bf p}-{\bf q})} \nonumber\\
&=&\int\!\frac{d {\bf q}}{(2\pi)^3} \:\lim_{\eta\downarrow 0}\frac{1}{2\pi i}\:\oint_C\:dz \frac{ e^{\eta z}}{e^{\beta z}-1}
\frac{1}{z -\xib ( {\bf q})}\frac{1}{i\omega-z -\xif ({\bf p}-{\bf q})} \nonumber\\
&=&\int\!\frac{d {\bf q}}{(2\pi)^3}\:\frac{1-n_f({\bf p}-{\bf q})
+n_b( {\bf q} )}{\xif({\bf p}-{\bf q})+\xib({\bf q})-i\omega} \: ,
\label{matsu_G}
\eeq 
\begin{figure}[t]
\begin{center}
\includegraphics[width=7cm]{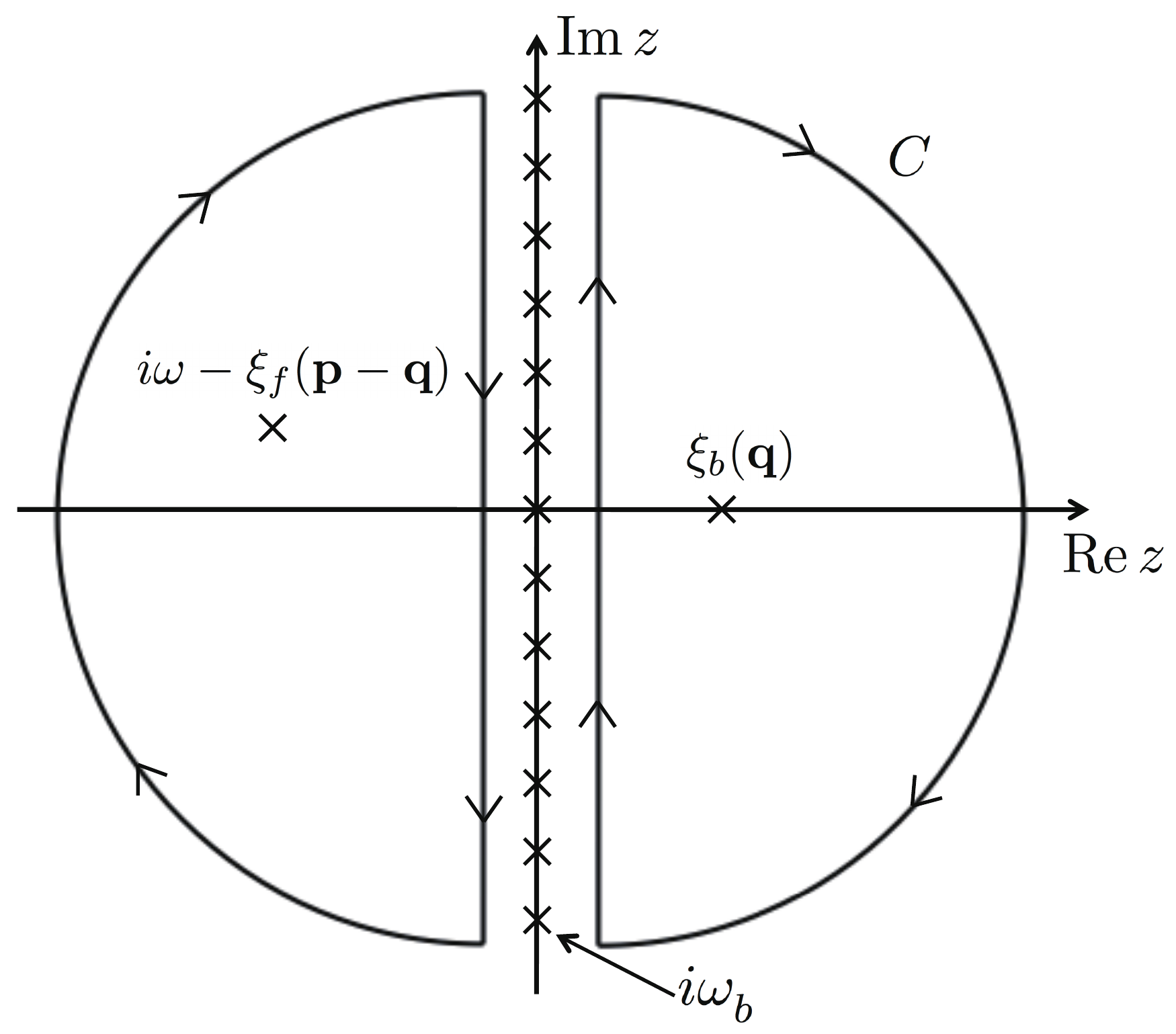}
\end{center}
\vspace{-0.5cm}
\caption{Contour for evaluation of the Matsubara frequency sum in Eq.(\ref{matsu_G}). 
The crosses indicate the position of the poles in the integrand. 
}
\label{fig_cont1}
\end{figure}
where $n_b( {\bf k} )$ and $n_f( {\bf k} )$ denote 
the Bose-Einstein and Fermi-Dirac distribution functions, respectively, 
\beq
 n_b( {\bf k} )&=& \frac{1}{e^{\beta\xib({\bf k})}-1} \:\: , \\
  n_f( {\bf k} )&=& \frac{1}{e^{\beta\xif({\bf k})}+1} \: ,
 \eeq
with $\xib$ and $\xif$ kinetic energies of single boson and fermion 
relative to the chemical potentials $\mub$ and $\muf$, respectively: 
$\xib(\bold{k})=\bold{k}^2/2\mb -\mub \: , \: 
\xif(\bold{k})=\bold{k}^2/2\mf-\muf $. 
Here we have taken a standard contour $C$ on the complex $z$-plane (see Fig.\:\ref{fig_cont1}) in order to 
convert the summation over the bosonic Matsubara frequency $\omb$ 
into a complex-integration along $C$ \cite{NO1988}. 
Then the inverse propagator Eq.(\ref{F_prop}) reads 
\beq
G^{-1}(p)
 &=&  \frac{\mR}{2\pi \abf} - \int\!\frac{d {\bf q}}{(2\pi)^3}\left\{ \frac{1}{ \varepsilon ( {\bf q} )}-
 \frac{1-n_f({\bf p}-{\bf q})+n_b({\bf q} )}{\xif({\bf p}-{\bf q})+\xib({\bf q})-i\omega}
 \right\}\: ,
 \label{F_prop2}
 \eeq
where the kinetic energy in the relative coordinate $\varepsilon ({\bf q})$ 
is defined by $\varepsilon ({\bf q})={\bf q}^2/(2\mR)$, 
and by using Eq.(\ref{swave}) 
we replaced the coupling constant $\gbf$ with the scattering length $\abf$. 
Note that up to this stage there is no need to put any assumption related to 
the strength of our coupling constant $\gbf$, or $n_{\rm tot}^{1/3}\abf$. 

Now let us study a strongly-coupled mixture characterized by $0\leq n_{\rm tot}^{1/3}\abf\ll1$. 
In this case, it is natural to consider the situation that both $\mub$ and $\muf$ are almost equal to $-\obf/2$, 
where $\obf$ is a binding energy of an isolated boson-fermion pair 
in the vacuum: $\obf=1/(2\mR a_{bf}^2)$. 
This is in accordance with the fact that 
in the strongly-coupled mixture 
the system becomes a dilute gas of CFs due to $n_{\rm tot}^{1/3}|\abf|\ll1$. 
We will later see that the number equations to relate 
the chemical potentials and the particle density indeed have a solution $\mub+\muf\simeq -\obf$. 
This implies putting one more pair reduces the total energy 
by an energy almost equal to $\obf$.

Then, the low energy and low momentum expansions of Eq.(\ref{F_prop2}) at zero temperature 
gives (see Appendix \ref{Cal_G}) 
\beq
G^{-1}({\bf p}, E)&\simeq&
\frac{\mR}{2\pi \abf} - \frac{(2\mR)^{3/2} }{4\pi } \sqrt{\:|\mu|+\frac{{\bf p}^2}{2(\mb+\mf)} -E\:}\: ,
\label{low_G}
\eeq
where $\mu$ denotes a total boson-fermion chemical potential: $\mu=\muf + \mub (< 0)$, 
and the chemical potentials are yet to be determined. 
We proceed to expand Eq.(\ref{low_G}) in terms of $\{{\bf p}^2/[2(\mb+\mf)] -E\}/|\mu|$, 
to obtain the derivative expansion of the inverse propagator, 
\beq
G^{-1}({\bf p},E)&\simeq& a\:-\: c\:{\bf p}^2 \:+\: d\: E
\:,\label{deriv_G}
\eeq
with the zero temperature coefficients 
\beq
a &=&\frac{\mR}{2\pi \abf}-\frac{\mR\sqrt{2\mR|\mu|}}{2\pi} \:,\label{coeff_a}\\
c&=&\frac{1}{4\pi}\frac{\mR}{\mf+\mb}\sqrt{\frac{\mR}{2|\mu|}} \:,\label{coeff_c}\\
d &=& \frac{\mR}{2\pi}\sqrt{\frac{\mR}{2|\mu|}} \:.\label{coeff_d}
\eeq
Thus in low-energy scales, Eq.(\ref{LO_p}) can be approximated by its effective action, 
\beq
{\cal S}_{\rm LO}\bigl[\bar{\cal F}_{\sigma},{\cal F}_{\sigma}\bigl]
 &\simeq&- \int\! dp \sum_{\sigma=\uparrow,\downarrow}\sqrt{d} \bar{\cal F}_\sigma(p)
\biggl(i\omega -\frac{{\bf p}^2}{2\bigl[d/(2c)\bigl]}+\frac{a}{d} \biggl) \sqrt{d}{\cal F}_\sigma(p)\: .
\label{LO_p2}
\eeq
Performing a proper normalization of CF fields with 
$\Psi_{\sigma}=\sqrt{d}{\cal F}_{\sigma}\: ,\: \bar{\Psi}_{\sigma}=\sqrt{d}\bar{\cal F}_{\sigma}$ 
yields a low-energy effective action, 
\beq
{\cal S}_{\rm LO}^{\rm eff.}\bigl[\bar{\Psi}_{\sigma},\Psi_{\sigma}\bigl]
&=&- \int\! dp  \sum_{\sigma=\uparrow,\downarrow} \bar{\Psi}_\sigma(p)
\biggl(i\omega -\frac{{\bf p}^2}{2\mF}+\muF \biggl)\Psi_\sigma(p)\: ,
\label{LO_p_eff}
\eeq
with the kinetic mass $ \mF=d/(2c)$ and 
the chemical potential $ \muF= a/d$ 
for the normalized CF fields $\Psi_{\sigma}, \bar{\Psi}_{\sigma}$: 
\beq
 \mF &=&\mf+\mb \:,  
 \label{mass_F}\\
 \muF &=& 2 \ombf \bigl( \sqrt{X}- X \bigl) \: .
  \label{chemipot_F}
\eeq
Here we have defined a dimensionless parameter $X = |\mu|/\ombf$. 
Using the effective action Eq.(\ref{LO_p_eff}), 
we can construct an effective theory described by the following partition function, 
\beq
Z_{\rm LO}^{\rm eff.}
&=&c \bigl[ Z_0^b(\mub) Z_0^f(\muf)\bigl]^N \int 
\biggl( \prod_{\sigma=\uparrow,\downarrow}{\cal D}\bar{\Psi}_{\sigma} {\cal D}\Psi_{\sigma}\biggl)
\exp \bigl\{- {\cal S}_{\rm LO}^{\rm eff.}\bigl[\bar{\Psi}_{\sigma},\Psi_{\sigma}\bigl] \bigl\}~,
\label{LO_Z_eff}
\eeq
which is valid for phenomena dominated by low-energy and low-momentum scales such that 
$\omega/|\mu|\ll 1,~\varepsilon({\bf p})/|\mu|\ll1$, and the chemical potential $|\mu|$ is determined 
by number equations,  
\beq
n_{\rm tot}&=&-\frac{\partial}{\partial \mub}\ln Z
~=~-\frac{\partial}{\partial \muf}\ln Z \: .
\label{num_eq}
\eeq

\begin{figure}[t]
\begin{center}
\includegraphics[width=7cm]{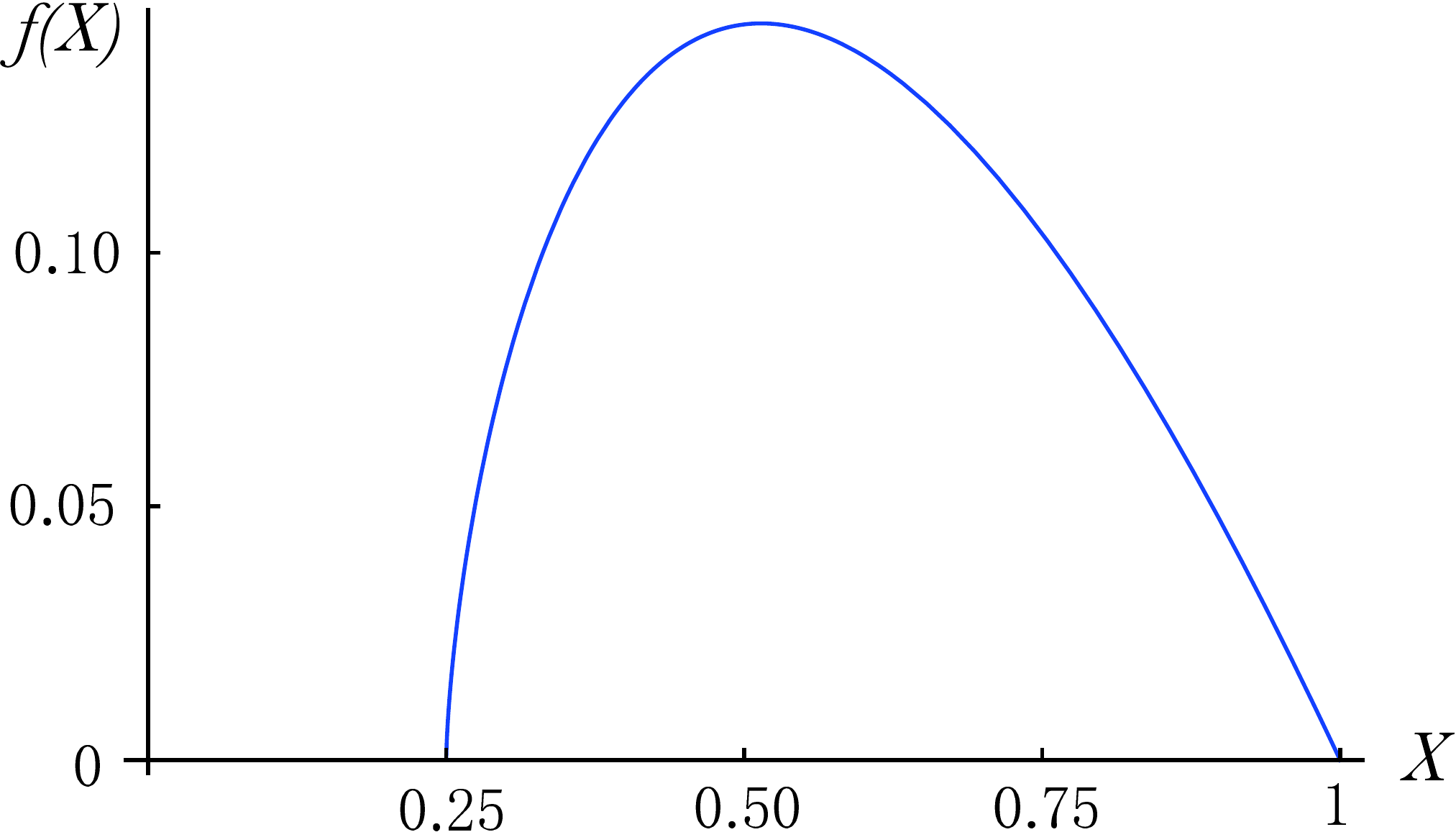}
\end{center}
\vspace{-0.5cm}
\caption{Numerical plot of $f(X)=(\sqrt{X}-X)(2-1/\sqrt{X})^{2/3}$ as a function of $X$. 
}
\label{fig_f_X}
\end{figure}

We will estimate chemical potentials, $\muf$, $\mub$ and $\muF$, at zero temperature, where 
the number equations reduce to  
\beq
n_{\rm tot}&=&-\frac{\partial}{\partial \mub}\ln Z
\nonumber\\
&=&-\frac{\partial}{\partial \mub}\ln Z_{\rm LO}^{\rm eff.}
\:-\:N\frac{\partial}{\partial \mub}\ln Z_0^b(\mub)
\nonumber\\
&=&-\biggl(\frac{\partial}{\partial \muF}\ln Z_{\rm LO}^{\rm eff.}\biggl) \times 
\biggl(\frac{\partial \muF}{\partial \mub}\biggl)
\:+\:N\int\!\!\frac{d {\bf k}}{(2\pi)^3} n_b({\bf k} )
\nonumber\\
&=&\frac{(2\mF\muF)^{3/2}}{3\pi^2} \times 
\biggl(2-\frac{1}{\sqrt{X}}\biggl) \: .
\label{eff_num_eq}
\eeq
Here we used the fact that $n_b({\bf k})$ vanishes at $T=0$ with $\mub<0$ 
and also that the effective action Eq.(\ref{LO_p_eff}) is the same action as for 
a two-component free Fermi gas with a mass $\mF$ and a chemical potential $\muF$. 
We can rewrite Eq.(\ref{eff_num_eq}) as a dimensionless equation, 
\beq
\frac{(3\pi^2)^{2/3}}{2}\biggl(\frac{\mR}{\mF}\biggl)\bigl(n_{\rm tot}^{1/3}\abf\bigl)^2
&=&\bigl(\sqrt{X}-X\bigl)\biggl(2-\frac{1}{\sqrt{X}}\biggl)^{2/3}\nonumber\\
&=:&f(X)\:,
\label{eff_num_eq2}
\eeq
where we defined a function $f$ as $f(X)=(\sqrt{X}-X)(2-1/\sqrt{X})^{2/3}$. 
Figure \ref{fig_f_X} shows a numerical plot of $f(X)$ as a function of $X$, and 
we can see that $f(X)$ becomes zero at $X=1/4$ and $X=1$. 
Since the left hand side of Eq.(\ref{eff_num_eq2}) becomes quite small in the strongly-coupled mixture, 
Eq.(\ref{eff_num_eq2}) will give two solutions around $X\sim 1/4$ and $X\sim 1$. 
From now on, we will focus on a solution $X\sim 1$ which is consistent with 
the  case of the dilute gas as we have mentioned before. 

Based on the above argument, 
we introduce a small positive parameter $\delta(=1-X)$ 
which should be determined through the number equations. 
Putting this into Eq.(\ref{eff_num_eq}) yields 
\beq
\frac{(3\pi^2 n_{\rm tot})^{2/3}}{2\mF}
&=&\muF \times 
\biggl(2-\frac{1}{\sqrt{X}}\biggl)^{2/3}
\nonumber\\
&\simeq&
\obf\delta~,
\eeq
which finally gives
\beq
|\mu|&=&\obf-\epsilon_F~+\: O(\epsilon_F/\obf)  \; ,
\label{mu_LO}\\
\muF&=&\epsilon_F~+\: O(\epsilon_F/\obf)   \; ,
\label{mF_LO}
\eeq
with a Fermi energy of CFs $\epsilon_F=(3\pi^2 n_{\rm tot})^{2/3}/(2\mF)$. 
Note that the above analysis becomes reliable 
only with a small $\delta(\simeq\epsilon_F/\obf)$ which demands the following condition:  
\beq
\frac{\epsilon_F}{\obf}&=&\frac{(3\pi^2)^{2/3}\; n_{\rm tot}^{2/3}}{2\mF}\times\biggl(\frac{1}{2\mR a_{bf}^2}\biggl)^{-1}
~=~(3\pi^2)^{2/3}\biggl(\frac{\mR}{\mF}\biggl)\bigl(n_{\rm tot}^{1/3}\abf\bigl)^2
~\ll~1~.  
\label{const_N}
\eeq
Now we can see that Eq.(\ref{const_N}) is automatically satisfied 
in our strongly-coupled mixture such that $0\leq n_{\rm tot}^{1/3}\abf\ll1$. 

Combining Eqs.(\ref{LO_p_eff}) and (\ref{mF_LO}), we find that 
in the leading order of the $1/N$-expansion 
a low-energy effective theory of the strongly-coupled boson-fermion mixture, 
whose energy scales satisfy $T/\obf\ll 1,~\omega/\obf\ll 1$ and $\varepsilon({\bf p})/\obf\ll1$, 
becomes just a two-component free Fermi gas of 
CFs with a mass $\mF=\mb+\mf$ and 
the same number density as the total number density of the original fermions $n_{\rm tot}$.
We remark that in our strongly-coupled mixture 
the small expansion parameters are $1/N$ and $n_{\rm tot}^{1/3}\abf$. 

\subsection{The next-to-leading-order term}

\begin{figure}[t]
\begin{center}
\includegraphics[width=14cm]{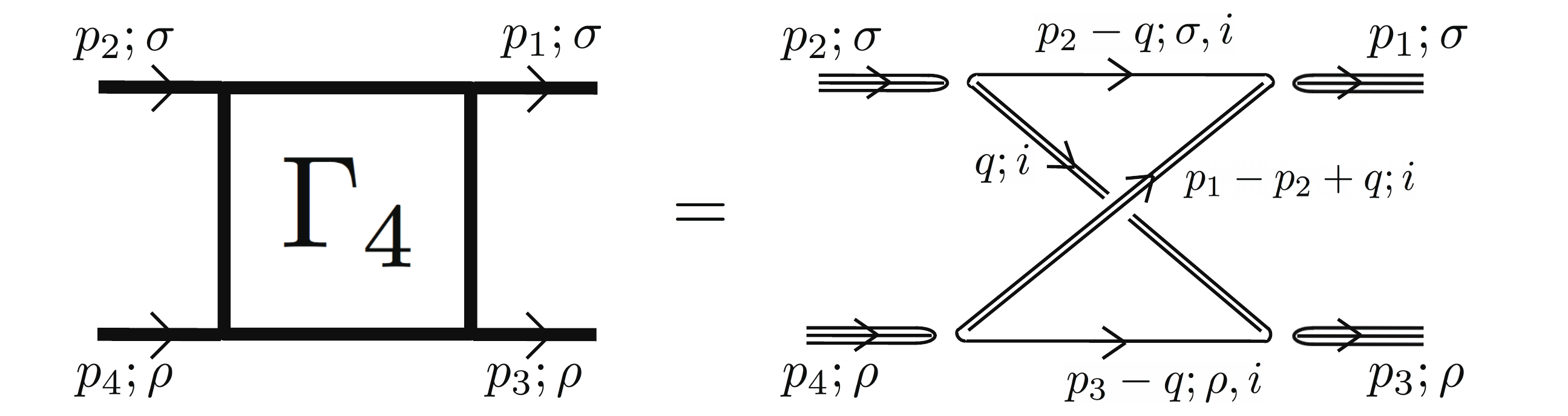}
\end{center}
\vspace{-0.5cm}
\caption{Graphical representation of the proper 4-point vertex of ``bare" CF fields 
$F_{\sigma}$ (again, not for ${\cal F}_{\sigma}$). 
In order to obtain the proper 4-point vertex of ${\cal F}_{\sigma}$, 
we need to multiply both sides by $(\sqrt{N})^4$, which comes from the normalization of four external legs 
(${\cal F}_{\sigma}= \sqrt{N}F_{\sigma}$). 
}
\label{fig3}
\end{figure}

We proceed to study the next-to-leading-order (NLO) term in Eq.(\ref{1/N_full_action}), 
and we will show that 
the NLO term gives an effective attraction between CFs in low-energy scales. 
The NLO term in Eq.(\ref{1/N_full_action}) becomes
\beq
 \lefteqn{{\cal S}_{\rm NLO}\bigl[\bar{\cal F}_{\sigma},{\cal F}_{\sigma}\bigl]}\nonumber\\
 &=&-\frac{1}{2N}\mathrm{tr}\int\! dvdwdz \sum_{\sigma,\rho=\uparrow,\downarrow} 
D(x,w){\cal A}_{\sigma}(w,v)D(v,z){\cal A}_{\rho}(z,y) \nonumber\\
&=&-\frac{1}{2N}\sum_{\sigma,\rho=\uparrow,\downarrow} 
\int\! dxdvdwdz\: D(x,w)\bar{{\cal F}}_\sigma(v)S(v,w){\cal F}_\sigma (w)D(v,z)\bar{{\cal F}_\rho}(x)S(x,z){\cal F}_{\rho}(z)
\nonumber\\
&=&-\frac{1}{2N} \sum_{\sigma, \rho=\uparrow,\downarrow} 
\int\! \biggl(\:\prod _{i=1} ^4 dp_i \biggl)\delta(p_4+p_2-p_1-p_3)
  \nonumber \\
 &&  \times \: 
 \biggl[\:\int\! dq\: D(q)S(p_2-q)D(p_1-p_2+q)S(p_3-q) \biggl]
 \bar{\cal F}_{\sigma}(p_1)\bar{\cal F}_{\rho}(p_3)
 {\cal F}_{\rho} (p_4){\cal F}_{\sigma} (p_2)
 \nonumber\\
 &=&
 \frac{1}{2}\sum_{\sigma, \rho=\uparrow,\downarrow} \int\! \biggl(\:\prod _{i=1} ^4 dp_i \biggl)\delta(p_4+p_2-p_1-p_3)
  \: \Gamma_4\bigl(\{ p_i \}_{i=1}^4\bigl)\bar{\cal F}_{\sigma}(p_1)\bar{\cal F}_{\rho}(p_3)
 {\cal F}_{\rho} (p_4){\cal F}_{\sigma} (p_2) \: ,
 \label{NLO_full}
\eeq
where $\Gamma_4$ represents the proper 4-point vertex of CFs, defined by 
\beq
 \Gamma_4\bigl(\{ p_i \}_{i=1}^4\bigl)&=&- \frac{1}{N}\int\! dq\: D(q)S(p_2-q)D(p_1-p_2+q)S(p_3-q) \:,
 \label{4v_full}
\eeq
with a set of momenta $\{ p_i \}_{i=1}^4=\{ p_1,\:p_2,\:p_3,\:p_4 \}$ constrained by 
the energy-momentum conservation, $p_4+p_2-p_1-p_3=0$. 
The graphical representation of Eq.(\ref{4v_full}) is shown
in Fig.\:\ref{fig3}, which indicates that the CFs interact  through
 the exchange of their constituent particles. 
We can see that the right hand side of Fig.\:\ref{fig3} 
gives large $N$ power-counting factors (i) $N$ from an internal loop, 
(ii) $(1/N)^4$ from four vertices, and (iii) $(\sqrt{N})^4$ from the normalization 
(${\cal F}_{\sigma}= \sqrt{N}F_{\sigma}$), 
which yield an $O(1/N)$ term as is $\Gamma_4$ in Eq.(\ref{4v_full}). 

For a low-energy effective theory, 
we expand the proper 4-point vertex in terms of $E/\ombf$ 
with a small energy scale $E$ relative to its binding energy $\obf$. 
Then the dominant contribution becomes 
\beq
\Gamma_4\bigl(\{ p_i \}\bigl) &=&
\Gamma_4 \bigl(\{ (\textbf{0},\pi T),(\textbf{0},\pi T),(\textbf{0},-\pi T),(\textbf{0},-\pi T) \} \bigl) 
\:\: + \:\: O(E/\ombf) \: . 
\label{4v_eff}
\eeq
We note that at finite temperatures 
it is impossible to put the frequencies in $\Gamma_4$ equal to zero, 
since the CF fields in Eq.(\ref{NLO_full}) are Grassmann fields (purely fermionic) 
and do not have Matsubara zero mode,
\textit{i.e.}, $\omega=(2n+1)\pi T$, $n\in \mathbb{Z}$. 
Note also that our procedure is essentially the same as 
in the derivation of the effective theory for Cooper pairs, 
which can be considered as composite bosons 
in two-component Fermi gases \cite{SRE93,HAU93,PS96,PS00,NS07,VSR07,AB08}. 
Let us denote the low-energy effective vertex in Eq.(\ref{4v_eff}) by $\Gamma_4(0)$, 
which becomes 
\beq
\Gamma_4(0) &=& \Gamma_4 \bigl(\{ (\textbf{0},\pi T),(\textbf{0},\pi T),
(\textbf{0},-\pi T),(\textbf{0},-\pi T) \} \bigl) \nonumber \\
 &=&\!\!-\frac{1}{N}\:T\sum_{\omb}\int\!\frac{d {\bf q}}{(2\pi)^3}
 D({\bf q},\omb)S(-{\bf q}, \pi T-\omb)D({\bf q},\omb)S(-{\bf q},-\pi T-\omb)\nonumber \\
 &=&\!\!-\frac{1}{N}\int\!\frac{d {\bf q}}{(2\pi)^3}\: T\sum_{\omb}
  \frac{1}{i\pi T-i\omb-\xif(-{\bf q})}\frac{1}{-i\pi T-i\omb-\xif(-{\bf q})}
  \left( \frac{1}{i\omb -\xib ( {\bf q})} \right)^2  . 
  \label{4v_eff2}
\eeq
We can perform the summation over the bosonic Matsubara frequency $\omb$ 
in Eq.(\ref{4v_eff2}) as follows, 
\beq
\lefteqn{T\sum_{\omb}
  \frac{1}{i\pi T-i\omb-\xif(-{\bf q})}\frac{1}{-i\pi T-i\omb-\xif(-{\bf q})}
  \left( \frac{1}{i\omb -\xib ( {\bf q})} \right)^2} \nonumber\\
 &=&\!\!
 \lim_{\eta\downarrow 0} \frac{1}{2\pi i}\:\oint_C\: dz\frac{e^{\eta z}}{e^{\beta z}-1}\:\frac{1}{z-i\pi T+\xif({\bf q})}
 \frac{1}{z+i\pi T+\xif({\bf q})}\:
  \frac{\partial }{\partial s}\left( \frac{1}{z-\xib ( {\bf q})-s} \right)\biggl|_{s=0} \nonumber \\
&=& \bigl[1-n_f( {\bf q})+n_b( {\bf q})\bigl]
 \frac{2\:\bigl[\xif( {\bf q})+\xib( {\bf q})\bigl]}{\bigl\{\bigl[\xif( {\bf q})+\xib( {\bf q})\bigl]^2+(\pi T)^2\bigl\}^2} 
 -\frac{1}{\bigl[\xif( {\bf q})+\xib( {\bf q})\bigl]^2+(\pi T)^2}
 \frac{\partial n_b( {\bf q})}{\partial \xib( {\bf q})}\:.
 \label{matsu_Gamma}
\eeq
\begin{figure}[t]
\begin{center}
\includegraphics[width=7cm]{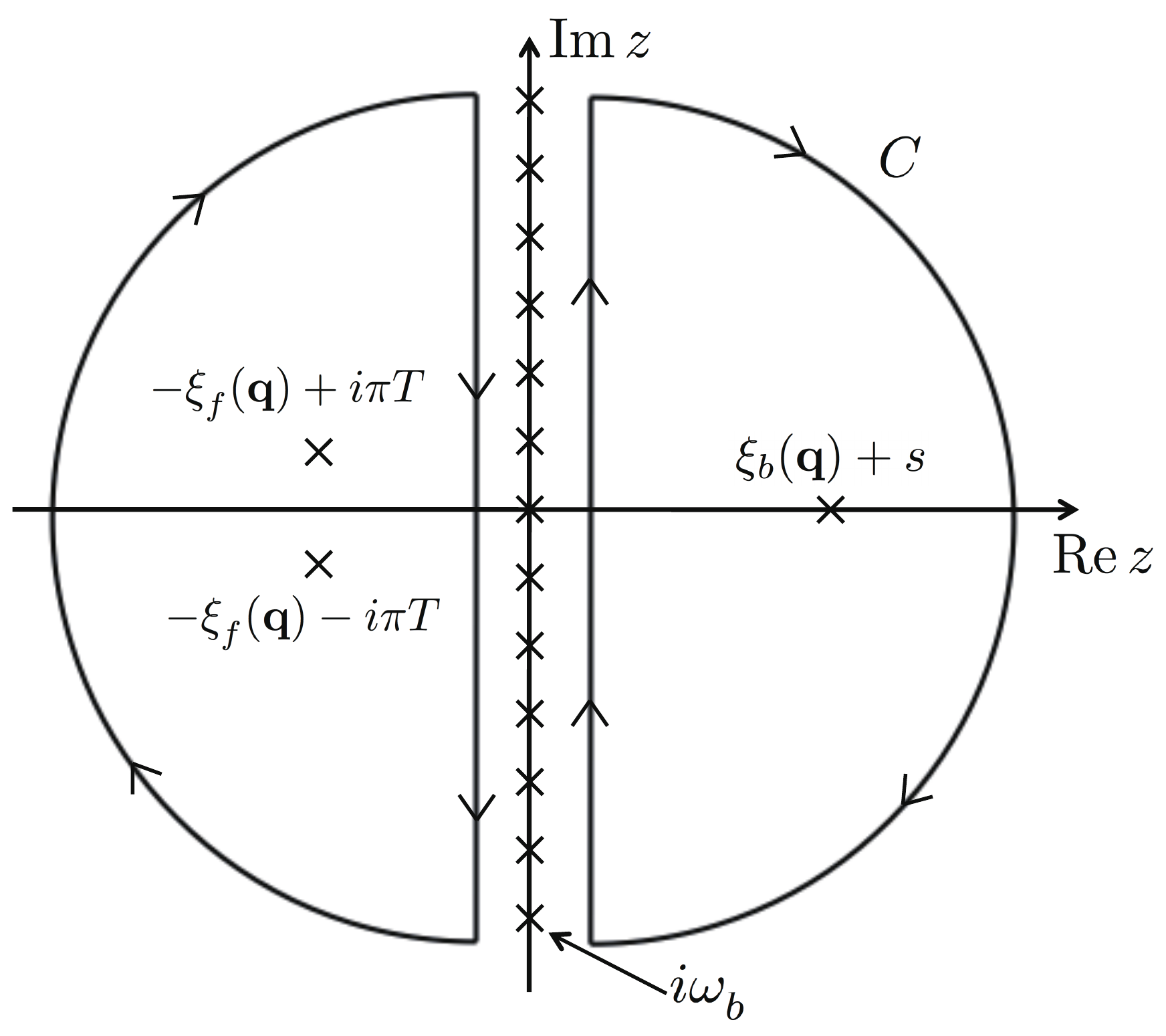}
\end{center}
\vspace{-0.5cm}
\caption{Contour for evaluation of the Matsubara frequency sum in Eq.(\ref{matsu_Gamma}). 
The crosses indicate the position of the poles in the integrand. 
}
\label{fig_cont2}
\end{figure}
Here as in the calculation of Eq.(\ref{matsu_G}), 
we have taken a standard contour $C$ on the complex $z$-plane (see Fig.\:\ref{fig_cont2})
in order to convert the Matsubara summation into a complex-integration along $C$. 
Substituting Eq.(\ref{matsu_Gamma}) into (\ref{4v_eff2}) yields 
\beq
\Gamma_4(0)
&=&-\frac{1}{N}\int\!\!\frac{d {\bf q}}{(2\pi)^3}
\frac{2\bigl[\xif( {\bf q})+\xib( {\bf q})\bigl]\bigl[1-n_f( {\bf q})+n_b( {\bf q})\bigl]}
{\bigl\{\bigl[\xif( {\bf q})+\xib( {\bf q})\bigl]^2+(\pi T)^2\bigl\}^2} 
\nonumber\\
&&
\:+\: \frac{1}{N}\int\!\!\frac{d {\bf q}}{(2\pi)^3}\frac{1}{\bigl[\xif( {\bf q})+\xib( {\bf q})\bigl]^2+(\pi T)^2}
\frac{\partial n_b( {\bf q})}{\partial \xib( {\bf q})} \:.
\eeq
We will estimate $\Gamma_4(0)$ analytically at $T=0$ with the same assumption as before, 
that is, with an assumption that both $\mub$ and $\muf$ are negative 
and their magnitudes are almost equal to $\obf/2$. 
This assumption still holds 
since the NLO term Eq.(\ref{NLO_full}) 
is suppressed by a factor $1/N$ compared to the LO terms 
and does not change the values of chemical potentials so much 
from Eq.(\ref{mu_LO}) and (\ref{mF_LO}).  
Under this assumption, $\Gamma_4(0)$ at $T=0$ reduces to 
\beq
\Gamma_4(0)
\!\!&=&\!\!
-\frac{1}{N}\int\!\!\frac{d {\bf q}}{(2\pi)^3}\frac{2}{\bigl[\xif( {\bf q})+\xib( {\bf q})\bigl]^3} 
\nonumber\\
\!\!&=&\!\! -\frac{1}{N}\frac{(2\mR)^{3/2}}{\pi^2}\int_0^{\Lambda/\sqrt{2\mR}}
\frac{x^2}{(x^2+|\mu|)^3}dx
\nonumber\\
\!\!&=& \!\!
-\frac{1}{N}\frac{(2\mR)^{3/2}}{\pi^2}\frac{1}{8}\biggl\{
\frac{1}{|\mu|^{3/2}}\tan^{-1}\biggl(\frac{\Lambda}{\sqrt{2\mR|\mu|}} \biggl)
+\frac{(\Lambda/\sqrt{2\mR})^3-|\mu|\Lambda/(2\mR)}{|\mu|\bigl[|\mu|+\Lambda^2/(2\mR)\bigl]^2}
\biggl\}
\nonumber\\
\!\!&\simeq& \!\!
-\frac{1}{N}\frac{1}{2\pi}\biggl(\frac{\mR}{2|\mu|} \biggl)^{3/2} \: ,
\eeq
where we have neglected $O(\sqrt{2\mR|\mu|}/\Lambda)$ corrections in the finial step. 
Thus in low-energy scales, Eq.(\ref{NLO_full}) can be approximated by
\beq
\lefteqn{
 {\cal S}_{\rm NLO}\bigl[\bar{\cal F}_{\sigma},{\cal F}_{\sigma}\bigl]}\nonumber\\
 &\simeq&
  \frac{1}{2}\sum_{\sigma, \rho=\uparrow,\downarrow} 
  \int\! \biggl(\:\prod _{i=1} ^4 dp_i \biggl)\delta(p_4+p_2-p_1-p_3)
 \Gamma_4(0) \bar{\cal F}_{\sigma}(p_1)\bar{\cal F}_{\rho}(p_3)
 {\cal F}_{\rho} (p_4){\cal F}_{\sigma} (p_2) \: .
 \label{NLO_p}
 \eeq
According to the analysis on the LO terms, 
we perform the same normalization as in Eq.(\ref{LO_p_eff}), 
$\Psi_{\sigma}=\sqrt{d}{\cal F}_{\sigma},\: \bar{\Psi}_{\sigma}=\sqrt{d}\bar{\cal F}_{\sigma}$, 
which yields a low-energy effective action in the NLO, 
\beq
\lefteqn{
 {\cal S}^{\rm eff.}_{\rm NLO}\bigl[\bar{\Psi}_{\sigma}, \Psi_{\sigma}\bigl]}\nonumber\\
 &=&
  \frac{1}{2}\sum_{\sigma, \rho=\uparrow,\downarrow}  \int\! \biggl(\:\prod _{i=1} ^4 dp_i \biggl) \delta(p_4+p_2-p_1-p_3) \;
  \gF \bar{\Psi}_{\sigma}(p_1)\bar{\Psi}_{\rho}(p_3)
 {\Psi}_{\rho} (p_4){\Psi}_{\sigma} (p_2) \: .
 \label{NLO_eff}
\eeq
Here we defined an effective four-Fermi coupling constant by $\gF=\Gamma_4(0)/d^2$, which becomes 
\beq
\gF &=& - \frac{1}{N}\frac{2\pi \abf}{\mR\sqrt{X}} \:.
\label{gFF1}
\eeq
We apply the LO result Eq.(\ref{mu_LO}), {\it i.e.}, $X\simeq 1-\varepsilon_F/\obf$, to the above Eq.(\ref{gFF1}),  
and finally obtain 
\beq
 \gF &\simeq&
  - \frac{1}{N}\frac{2\pi \abf}{\mR}\biggl(1+\frac12\frac{\varepsilon_F}{\obf}\biggl)
 \nonumber\\
&\simeq&  - \frac{1}{N}\frac{2\pi \abf}{\mR} \:.
\label{gFF2}
\eeq
Introducing the density of states per unit volume at the Fermi surface $N_F(0)$: 
\beq
N_F(0)&=&\frac{\mF (3\pi^2 n_{\rm tot})^{1/3}}{\pi^2}~, 
\eeq
we have the dimensionless parameter 
for the strength of the effective four-Fermi interaction, 
\beq
N_F(0)\gF&\simeq&-\frac{\mF (3\pi^2 n_{\rm tot})^{1/3}}{\pi^2}\frac{2\pi\abf}{N \mR}
\nonumber\\
&=& -2\biggl(\frac{3}{\pi}\biggl)^{1/3}\frac{\mF}{\mR}\frac{n_{\rm tot}^{1/3}\abf}{N} \: . 
\label{4_fermi_str}
\eeq
Here we find that the NLO term yields an effective four-Fermi interaction, 
which is attractive and weak in a twofold meaning. 
First, we are considering strongly-coupled boson-fermion mixtures so that 
the dimensionless parameter $n_{\rm tot}^{1/3}\abf$ is positive and much smaller than 1, 
which makes $N_F(0)\gF$ negative and much smaller than 1, as discussed in Ref.\cite{MBH09} for $N=1$. 
Secondly, we also have large $N$ degrees of freedom in the original boson-fermion mixture, 
which yield the large suppression factor $1/N$ in Eq.(\ref{4_fermi_str}).

\subsection{BCS superfluidity of composite fermions}
From the results of the previous sections, we have a low-energy effective action 
up to the NLO term in the $1/N$-expansion: 
\beq
\lefteqn{{\cal S}^{\rm eff.}_{\rm LO}\bigl[\bar{\Psi}_{\sigma}, \Psi_{\sigma}\bigl]
\:+\:{\cal S}^{\rm eff.}_{\rm NLO}\bigl[\bar{\Psi}_{\sigma}, \Psi_{\sigma}\bigl]}\nonumber\\
 &=&
-\sum_{\sigma=\uparrow,\downarrow} \int\! dp   \: \bar{\Psi}_\sigma(p)
\biggl(i\omega -\frac{{\bf p}^2}{2\mF}+\muF \biggl)\Psi_\sigma(p)
 \nonumber\\
 &&\:+\:
  \frac{1}{2}\sum_{\sigma, \rho=\uparrow,\downarrow} 
   \int\! \biggl(\:\prod _{i=1} ^4 dp_i \biggl) \delta(p_4+p_2-p_1-p_3) \;
  \gF \bar{\Psi}_{\sigma}(p_1)\bar{\Psi}_{\rho}(p_3)
 {\Psi}_{\rho} (p_4){\Psi}_{\sigma} (p_2) \: ,
 \label{LO_NLO_eff}
\eeq
with the physical parameters of CF fields, 
\beq
 \mF &=&\mf+\mb \:,  \\
\muF&\simeq&\epsilon_F~=~\frac{(3\pi^2 n_{\rm tot})^{2/3}}{2\mF} \:,  \\
 \gF &\simeq&- \frac{1}{N}\frac{2\pi \abf}{\mR} \:.
\eeq
Equation (\ref{LO_NLO_eff}) is nothing but an action of two-component Fermi gases 
with weakly attractive four-Fermi interactions, which yields the BCS-paired state of fermions at low temperature. 
Thus we can expect that our system described by Eq.(\ref{LO_NLO_eff}) 
also favor the BCS superfluidity of CFs (CF-BCS) below a transition temperature \cite{BCS57,GMB61}, 
\beq
T_{\mathrm{C}}( \mathrm{CF}\mathchar`- \mathrm{BCS})\:=\:
\frac{\gamma}{\pi}\biggl(\frac{2}{e}\biggl)^{7/3}
\epsilon _F 
\exp\biggl(\frac{\pi}{\:2k_F \aF}  \biggl) \:,
\label{F_BCS}
\eeq
with the Fermi momentum $k_F=\sqrt{2\mF\epsilon _F}=(3\pi^2 n_{\rm tot})^{1/3}$ and
the s-wave scattering length given by
\beq
\aF&=&\frac{\mF}{4\pi}\gF
\nonumber\\
&=&-\frac{1}{N}\frac{\mF}{2\mR}\abf\: .
\label{FF_scat_len}
\eeq
Setting $N=1$ yields the same result as derived in our previous work \cite{MBH09}. 
Note that the coefficient of $\abf$ in Eq.(\ref{FF_scat_len}) becomes $-2$ for $N=1$ with $\mb = \mf$, 
which is the same in magnitude but opposite in sign from the scattering length between 
bosonic dimers composed of spin-singlet fermion pairs within the same approximation. 
This is because our CFs are different in the statistics of their constituent particles 
from the composite bosons, so-called Cooper pairs, 
in two-component Fermi gases \cite{SRE93,HAU93,KBEK04}.

\subsection{Higher order terms in the $1/N$-expansion}
 
We now consider the higher order terms in the $1/N$-expansion 
of the CF action Eq.(\ref{1/N_full_action}), 
especially focusing on the corrections to the effective four-Fermi. 
 
\begin{figure}
\begin{center}
\includegraphics[width=10cm]{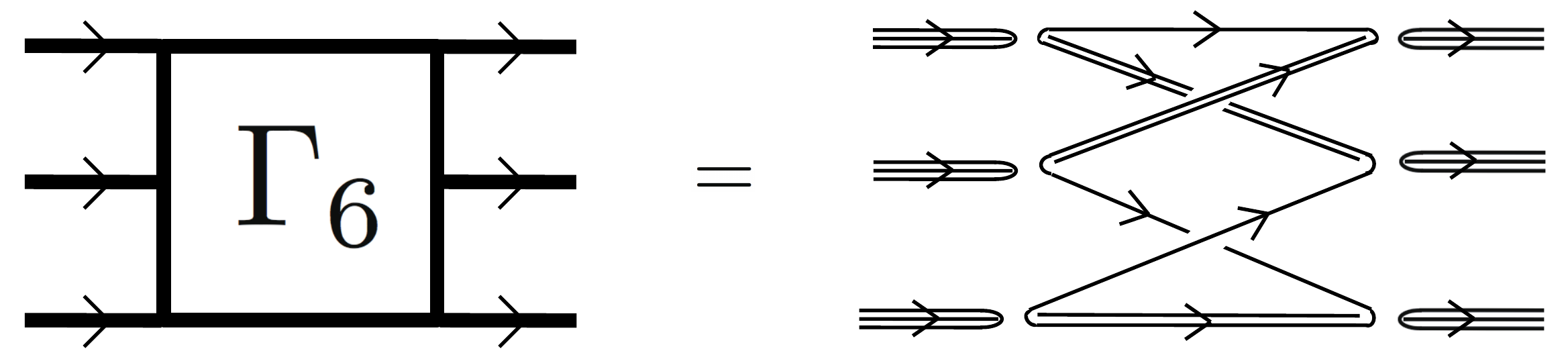}
\end{center}
\vspace{-0.5cm}
\caption{Graphical representation of the proper 6-point vertex of ``bare" CF fields 
$F_{\sigma}$. In order to obtain the proper 6-point vertex of ${\cal F}_{\sigma}$, 
we need to multiply both sides by $(\sqrt{N})^6$, which comes from the normalization of six external legs 
(${\cal F}_{\sigma}= \sqrt{N}F_{\sigma}$). 
}
\label{fig4}
\end{figure}

Figure \ref{fig4} shows a graphical representation of the proper 6-point vertex, 
based on the sextet term of CF fields in Eq.(\ref{1/N_full_action}). 
We can see that the right hand side of Fig.\:\ref{fig4} gives large $N$ power-counting factors 
(i) $N$ from an internal loop, (ii) $(1/N)^6$ from six vertices, 
and (iii) $(\sqrt{N})^6$ from the normalization (${\cal F}_{\sigma}= \sqrt{N}F_{\sigma}$), 
which yield an $O(1/N^2)$ term as it should be in Eq.(\ref{1/N_full_action}). 

In general, we denote an $O(1/N^{n-1})$ term in Eq.(\ref{1/N_full_action}) by 
${\cal S}^{(n-1)}$ ($n\geq 2$), which is composed of $2n$ CF fields 
and the proper $2n$-point vertex, 
\beq
{\cal S}^{(n-1)}\bigl[\bar{\cal F}_{\sigma},{\cal F}_{\sigma}\bigl]
&=&
-\biggl(\frac{1}{N}\biggl)^{n-1}\frac{1}{n}
\tr \: \biggl[\int\! dw \sum_{\sigma=\uparrow,\downarrow} D(x,w){\cal A}_{\sigma}(w,y)\biggl]^n
\nonumber\\
&=&
\frac{1}{n}\sum_{\sigma_1 , \dots , \sigma_n=\uparrow,\downarrow}
\int\biggl(\:\prod_{j=1}^{n} dp_{2j-1}dp_{2j}\biggl)
\:\delta\biggl( \:\sum_{j=1}^{n}p_{2j}-\sum_{j=1}^{n}p_{2j-1}\biggl)
\nonumber\\
&&\qquad\qquad \times\:
\Gamma_{2n}\bigl(\{ p_i \}_{i=1}^{2n}\bigl)
\biggl\{ \:\prod_{j=1}^{n}  \bar{\cal F}_{\sigma_j}(p_{2j-1}) {\cal F}_{\sigma_j} (p_{2j})  \biggl\} \:.
\label{O_(n-1)}
\eeq
Here $\Gamma_{2n}\bigl(\{ p_i \}_{i=1}^{2n}\bigl)$ represents 
a proper $2n$-point vertex of CF fields, 
defined by 
\beq
\Gamma_{2n}\bigl(\{ p_i \}_{i=1}^{2n}\bigl)\!\!\!\!&=&\!\!\!\!
-\frac{1}{N^{n-1}}
\int dq \prod_{j=1}^{n}  D\biggl(\:\sum_{k=1}^{j-1}p_{2k-1}-\sum_{k=1}^{j-1}p_{2k}+q\biggl)\: 
S\biggl(\:\sum_{k=1}^{j}p_{2k} -\sum_{k=1}^{j-1}p_{2k-1}-q\biggl) \:,
\label{2nv_full}
\eeq  
\begin{figure}
\begin{center}
\includegraphics[width=10cm]{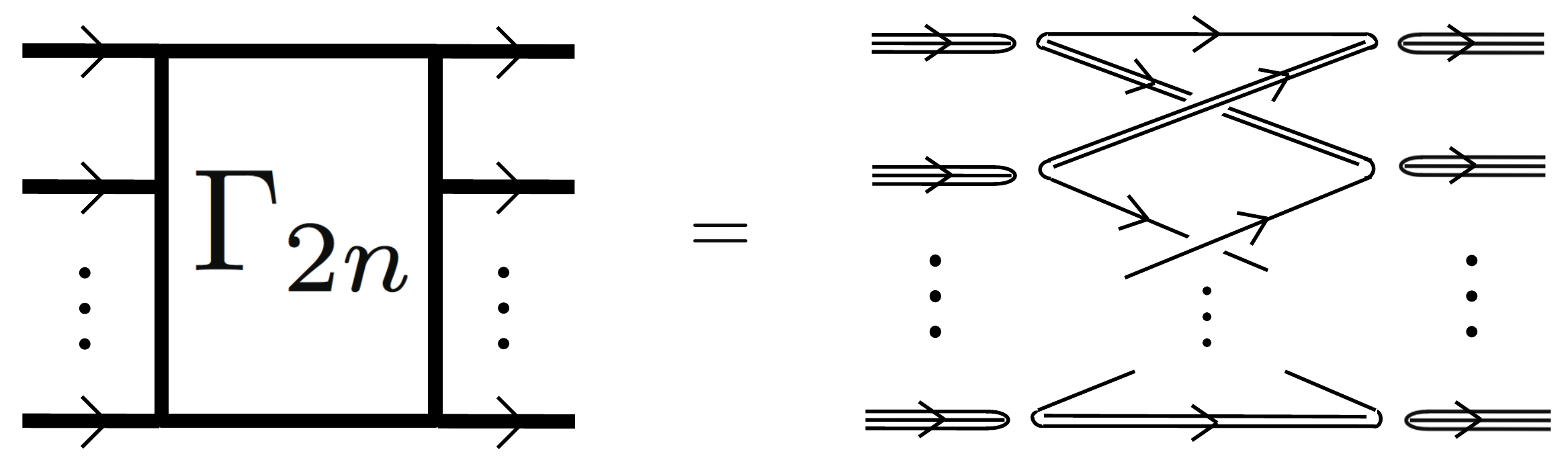}
\end{center}
\vspace{-0.5cm}
\caption{Graphical representation of the proper $2n$-points vertex of ``bare" CFs 
$F_{\sigma}$. 
In order to obtain the proper $2n$-points vertex of ${\cal F}_{\sigma}$, 
we need to multiply both sides by $(\sqrt{N})^{2n}$, which comes from the normalization of $2n$ external legs 
(${\cal F}_{\sigma}= \sqrt{N}F_{\sigma}$). 
}
\label{fig5}
\end{figure}
with a set of momenta $\{p_i\}_{i=1}^{2n}=\{p_1, p_2 , \dots , p_{2n}\}$ constrained by 
the energy-momentum conservation, $\sum_{j=1}^{n}p_{2j}-\sum_{j=1}^{n}p_{2j-1}=0$. 
Appendix \ref{FT} gives a detail derivation of the above expressions. 
Figure \ref{fig5} shows a graphical representation of the proper $2n$-point vertex. 
We can see that the right hand side of Fig.\:\ref{fig5} gives large $N$ power-counting factors 
(i) $N$ from an internal loop, (ii) $(1/N)^{2n}$ from $2n$ vertices, 
and (iii) $(\sqrt{N})^{2n}$ from the normalization (${\cal F}_{\sigma}= \sqrt{N}F_{\sigma}$), 
which yield an $O(1/N^{n-1})$ term as is $\Gamma_{2n}$ in Eq.(\ref{2nv_full}). 
Comparing Eqs.(\ref{O_(n-1)}), (\ref{2nv_full}) with Eqs.(\ref{NLO_full}), (\ref{4v_full}), 
we can easily check the consistency for the case of $n=2$, {\it i.e.}, 
${\cal S}^{(1)}\bigl[\bar{\cal F}_{\sigma},{\cal F}_{\sigma}\bigl]
={\cal S}_{\rm NLO}\bigl[\bar{\cal F}_{\sigma},{\cal F}_{\sigma}\bigl]$. 

\begin{figure}[b]
\begin{center}
\includegraphics[width=10cm]{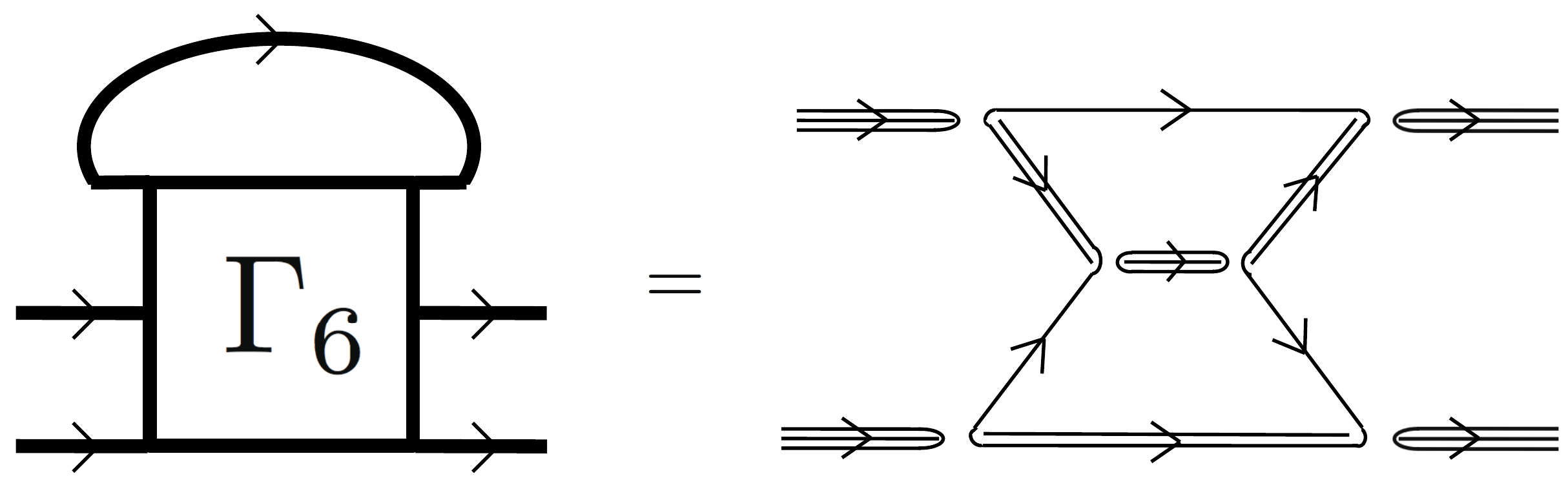}
\end{center}
\vspace{-0.5cm}
\caption{Graphical representation of the contraction of the proper 6-point vertex 
of ``bare" CF fields $F_{\sigma}$, 
which contributes to the effective four-Fermi interaction. 
}
\label{fig6}
\end{figure}

\begin{figure}
\begin{center}
\includegraphics[width=10cm]{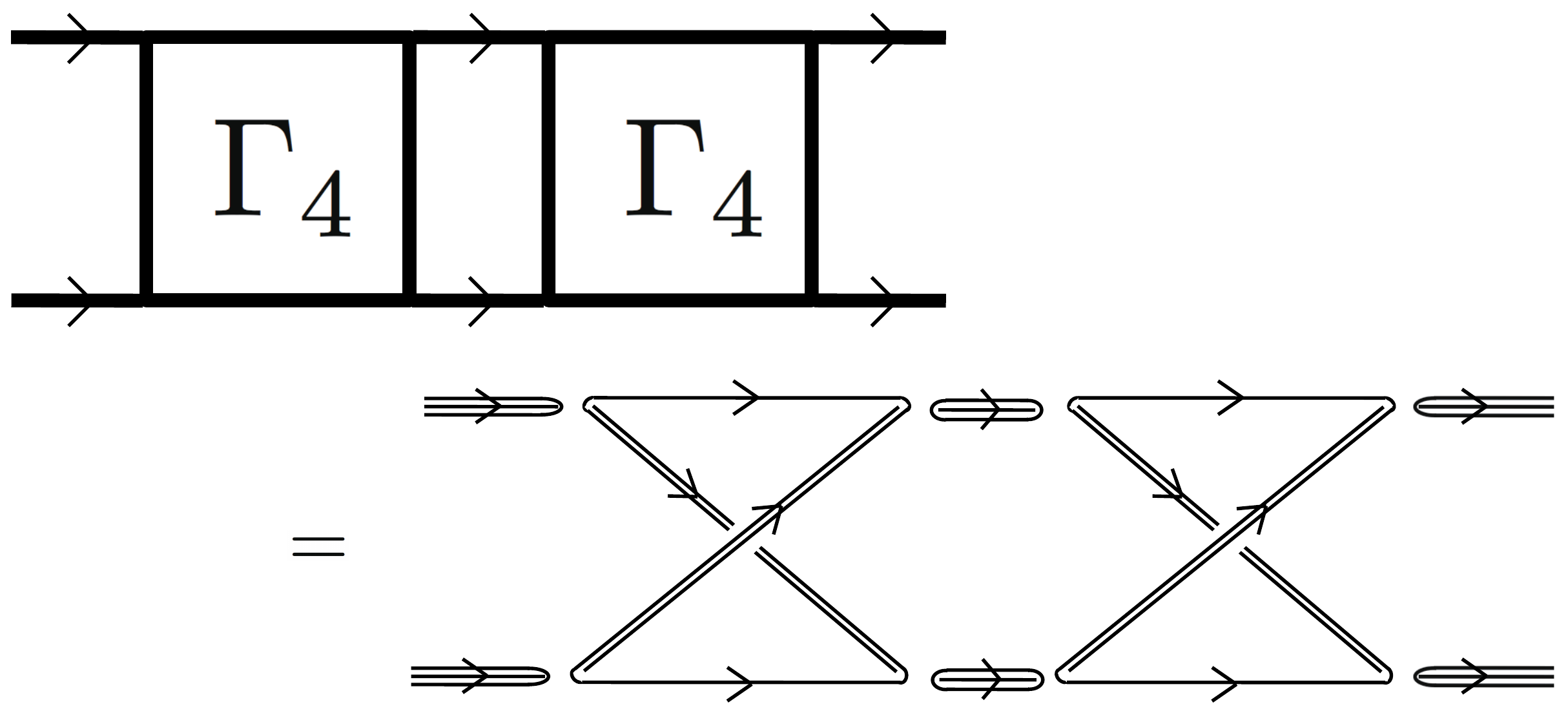}
\end{center}
\vspace{-0.5cm}
\caption{Graphical representation of the  shortest ladder diagram 
of the proper 4-point vertices of ``bare" CF fields $F_{\sigma}$, 
which contributes to the effective four-Fermi interaction. 
}
\label{fig7}
\end{figure}

Now let us consider diagrams higher order in the $1/N$-expansion 
which contribute to the effective four-Fermi interaction. 
The leading-order contribution to the effective four-Fermi interaction is given by the proper 4-point vertex $\Gamma_{4}$. 
Since our $1/N$-expansion is equivalent to the loop expansion in terms of CF fields, 
the corrections to the proper 4-point vertex $\Gamma_{4}$ should start from 
one-loop graphs, as shown in Fig.\:\ref{fig6} and \ref{fig7}. 
Figure \ref{fig6} represents the contraction of the proper 6-point vertex $\Gamma_{6}$, 
and its right hand side gives large $N$ power-counting
factors (i) $N$ from an internal loop, 
(ii) $(1/N)^{6}$ from six vertices, and (iii) $(\sqrt{N})^{6}$ from the normalization 
(${\cal F}_{\sigma}= \sqrt{N}F_{\sigma}$), 
which give an $O(1/N^{2})$ term in total, 
{\it i.e.}, the next-to-leading-order contribution 
to the effective four-Fermi interaction. 
Figure \ref{fig7} shows the  shortest ladder diagram 
of the proper 4-point vertices $\Gamma_{4}$, 
and its right hand side gives large $N$ power-counting
factors (i) $N^2$ from two internal loops, 
(ii) $(1/N)^{8}$ from eight vertices, and (iii) $(\sqrt{N})^{8}$ from the normalization 
(${\cal F}_{\sigma}= \sqrt{N}F_{\sigma}$), 
which again yield an $O(1/N^{2})$ term. 

It is also possible to consider general $O(1/N^{n})$ corrections based on the loop expansion 
in terms of CF fields which contribute to the effective four-Fermi interaction. 
Instead, we just remark that all the graphs are in the same order in terms of $n^{1/3}\abf$, 
thus for $N=1$ we need to sum up them to obtain the effective vertex function, 
as performed numerically in Refs\cite{PSS04,BKKCL06,LG06}. 

\section{Summary and discussion}

We have investigated the large $N$ expansion for strongly-coupled boson-fermion mixtures, 
proposed in Ref.\cite{MBH09}. 
We first derived a theory equivalent to the original boson-fermion mixture, 
which is described by composite fermions (CFs). 
The $1/N$-expansion naturally appears 
in the quantum theory of CFs. 
We showed that the leading-order terms in the $1/N$-expansion yield 
a low energy effective action of CFs 
which is equivalent to that of a two-component free Fermi gas. 
The next-to-leading-order term was also estimated, 
and it turned out that the effective action up to the NLO reduces to 
an action of a weakly-interacting two-component Fermi gas. 
Thus we concluded that there is the BCS superfluidity of CFs 
below $T_{\mathrm{C}}( \mathrm{CF}\mathchar`- \mathrm{BCS})$ given by Eq.(\ref{F_BCS}) 
in our large $N$ model. 
Also we discussed how to estimate the higher order terms in the $1/N$-expansion, 
where the diagrammatic representation provides simple explanations for power-counting 
of the $1/N$ factors. 

Finally, we would like to mention important similarities 
between our boson-fermion mixtures and hadron physics. 
Table \ref{UCA_QCD} summarizes the correspondence in components 
between ultracold atoms and dense QCD, 
both of which can be considered as boson-fermion mixtures with $N=1$ in our model \cite{BHTY08, MBH09, HM09}. 
Note that it is known that the effective interaction between nucleons is not so strong 
as the original gluonic interaction between quarks. 
For example, the energy gap in nuclear matter, {\it i.e.}, superfluid matter of nucleons, 
is at most a few MeV \cite{DHJ03}, 
while the energy gap in color-superconductivity is from 10 to $10^2$ times larger \cite{ASRS08}.  
This is consistent with our results which show that the weakly-coupled CF system 
can be derived from the strongly-coupled boson-fermion mixture. 
However, our model seems too simple to relate its results to various phenomena in QCD. 
Furthermore, chiral symmetry breaking plays an important role in hadron physics \cite{BHTY08}, 
which does not appear in nonrelativistic systems. 
Keeping these observations in mind, we suggest that 
both theoretical and experimental studies in boson-fermion mixtures 
provide a new tool to investigate properties of dense QCD, 
which is not readily observable in laboratory experiments. 
\begin{table}[t]
\caption{Correspondence between the boson-fermion mixture in ultracold atoms
and the diquark-quark mixture in dense QCD.}
\begin{center}
\begin{tabular}{|c||c|c|}
\hline
Our Notations & Ultra-Cold Atoms   & Dense QCD \\
\hline \hline
$\phi$ & bosonic atom (e.g.,$^{87}$Rb)  & diquark \\ \hline
$\psi_{\uparrow},\: \psi_{\downarrow}$ & fermionic atoms (e.g.,$^{40}$K) & unpaired quarks \\ \hline
$\Psi_{\uparrow},\: \Psi_{\downarrow}$ & composite fermions (boson-fermion dimers) 
&  nucleons (baryons) \\ \hline
$\gbf$ & boson-fermion attraction & gluonic attraction  \\ \hline
$\braket{\Psi_{\uparrow}\Psi_{\downarrow}}\neq0$
& composite-fermion superfluidity & nucleon superfluidity  \\ \hline
\end{tabular}
\end{center}
\label{UCA_QCD}
\end{table}

\section*{Acknowledgements}

The author thanks G.~Baym, D.~Blaschke, T.~Hatsuda, S.~Uchino for fruitful discussions. 
This research was supported by JSPS Research Fellowship for young scientists.


\appendix
\section{Fourier transformations}
\label{FT}

The definitions  of Fourier transforms used in our main text are given by 
\beq
{\cal F}_\sigma(x) &=&\int\! dp \:e^{ipx}{\cal F}_\sigma (p) \; ,\\
\bar{{\cal F}}_\sigma(x) &=& \int\! dp \:e^{-ipx}\bar{{\cal F}}_\sigma(p) \: ,\\
D(x,y) &=& \int\! dq \: e^{iq(x-y)}D(q) \; , \\ 
S(x,y) &=& \int\! dq \: e^{iq(x-y)}S(q) \: , 
\eeq
with an inner product: $ipx=i\omega\tau-i{\bold p}\cdot{\bold x}$ where 
$\omega$ and ${\bold p}$ denote the Matsubara frequency and spatial momentum vector, respectively. 
For simplicity, we have used notations: $p=(\textbf{p},i\omega)$, 
$\int\! dp= T\sum_{\omega}\omega \int\! d\textbf{p}/(2\pi)^3$. 
We adopt a convention which distinguishes functions 
from their Fourier transforms only by their arguments. 
Correspondingly, the Fourier transform of the quadratic term in Eq.(\ref{LO_x}) becomes  
\beq
\lefteqn{\int\! dxdw \:\bar{{\cal F}}_\sigma(x)D(x,w)S(x,w){\cal F}_\sigma (w) } \nonumber\\
 &=& \int\! dp_1 dp_2 dq_1 dq_2 \:\delta (q_1+q_2-p_1) \delta (p_2-q_1-q_2) \bar{{\cal F}}_\sigma(p_1) D(q_1) S(q_2) 
 {\cal F}_\sigma (p_2) \nonumber \\
 &=& \int\! dp\: \bar{{\cal F}}_\sigma(p) \biggl[ \:\int\! dq D(q) S(p-q) \biggl] {\cal F}_\sigma (p) \:.
\eeq
Also, the quartic term in Eq.(\ref{NLO_full}) transforms as 
\beq
\lefteqn{
-\frac{1}{N}\int\! dxdvdwdz 
D(x,w)\bar{{\cal F}}_\sigma(v)S(v,w){\cal F}_\sigma (w)D(v,z)\bar{{\cal F}_\rho}(x)S(x,z){\cal F}_{\rho}(z) }
\nonumber \\
 &=& -\frac{1}{N}\int\! \biggl( \prod _{i=1} ^4 dp_i dq_i \biggl) 
 \delta (q_1+q_4-p_3) \delta (p_2-q_1-q_2) \delta (q_2+q_3-p_1) 
 \delta (p_4-q_3-q_4) \nonumber \\
 & & \mspace{200mu}\times \: D(q_1)\bar{{\cal F}}_\sigma(p_1) S(q_2) {\cal F}_\sigma (p_2) D(q_3) \bar{{\cal F}_\rho}(p_3) 
 S(q_4) {\cal F}_\rho (p_4) \nonumber \\
 &=&\int\! \biggl( \prod _{i=1} ^4 dp_i \biggl) \delta (p_4+p_2-p_1-p_3)
\Gamma\bigl(\{p_i\}_{i=1}^4\bigl)\bar{{\cal F}}_\sigma(p_1) \bar{{\cal F}_\rho}(p_3) {\cal F}_\rho (p_4){\cal F}_\sigma (p_2)
\: ,
\eeq
where $\Gamma\bigl(\{p_i\}_{i=1}^4\bigl)$ represents a proper 4-point vertex of composite fermions, defined by 
\beq
\Gamma\bigl(\{p_i\}_{i=1}^4\bigl)&=&-\frac{1}{N}\int\!\! dq\: D(q)S(p_2-q)D(p_1-p_2+q)S(p_3-q) \: ,
\eeq  
with a set of momenta $\{p_i\}_{i=1}^4=\{p_1, p_2 , p_3 , p_4\}$. 

We can write down a general $2n$-point vertex function explicitly in its Fourier transform,  
\beq
\lefteqn{
-\frac{1}{N^{n-1}}
\tr \: \biggl[\int\! dw \sum_{\sigma=\uparrow,\downarrow} D(x,w){\cal A}_{\sigma}(w,y)\biggl]^n
}\\
&=&\!\!\!\!
-\frac{1}{N^{n-1}}\sum_{\sigma_1 , \dots , \sigma_n=\uparrow,\downarrow}
\int\biggl(\:\prod_{i=1}^{2n+1}dx_i \biggl)\delta(x_1-x_{2n+1})
\biggl\{\:\prod_{j=1}^{n} D(x_{2j-1},x_{2j}) \bar{\cal F}_{\sigma_j}(x_{2j+1})S(x_{2j+1},x_{2j}){\cal F}_{\sigma_j} (x_{2j}) 
\biggl\}
\nonumber\\
&=&\!\!\!\!
-\frac{1}{N^{n-1}}\sum_{\sigma_1 , \dots , \sigma_n=\uparrow,\downarrow}
\int\biggl(\:\prod_{j=1}^{n}dq_{2j-1}dq_{2j}dp_{2j-1}dp_{2j}\biggl)
\nonumber\\
&&
\qquad \qquad
\delta(q_1+q_{2n}-p_{2n-1})
\biggl\{ \:\prod_{j=1}^{n} \delta(p_{2j}-q_{2j}-q_{2j-1}) \biggl\}
\biggl\{ \:\prod_{j=1}^{n-1} \delta(q_{2j+1}+q_{2j}-p_{2j-1}) \biggl\}
\nonumber\\
&&\qquad \qquad  \times
\biggl\{ \:\prod_{j=1}^{n}  D(q_{2j-1}) S(q_{2j})  \biggl\}
\biggl\{ \:\prod_{j=1}^{n}  \bar{\cal F}_{\sigma_j}(p_{2j-1}) {\cal F}_{\sigma_j} (p_{2j})  \biggl\}
\nonumber\\
&=&\!\!\!\!
-\frac{1}{N^{n-1}}
\sum_{\sigma_1 , \dots , \sigma_n=\uparrow,\downarrow}
\int\biggl(\:\prod_{j=1}^{n} dp_{2j-1}dp_{2j}\biggl)
\:\delta\biggl( \:\sum_{j=1}^{n}p_{2j}-\sum_{j=1}^{n}p_{2j-1}\biggl)
\nonumber\\
&\times&\!\!\!
\biggl\{ \int dq_1\prod_{j=1}^{n}  D\biggl(\:\sum_{k=1}^{j-1}p_{2k-1}-\sum_{k=1}^{j-1}p_{2k}+q_1\biggl)\: 
S\biggl(\:\sum_{k=1}^{j}p_{2k} -\sum_{k=1}^{j-1}p_{2k-1}-q_{1}\biggl) 
 \biggl\}
\biggl\{ \:\prod_{j=1}^{n}  \bar{\cal F}_{\sigma_j}(p_{2j-1}) {\cal F}_{\sigma_j} (p_{2j})  \biggl\}
\nonumber\\
&=&\!
\sum_{\sigma_1 , \dots , \sigma_n=\uparrow,\downarrow}
\int\biggl(\:\prod_{j=1}^{n} dp_{2j-1}dp_{2j}\biggl)
\:\delta\biggl( \:\sum_{j=1}^{n}p_{2j}-\sum_{j=1}^{n}p_{2j-1}\biggl)
\Gamma_{2n}\bigl(\{ p_i \}_{i=1}^{2n}\bigl)
\biggl\{ \:\prod_{j=1}^{n}  \bar{\cal F}_{\sigma_j}(p_{2j-1}) {\cal F}_{\sigma_j} (p_{2j})  \biggl\} \:,
\nonumber
\eeq
where $\Gamma_{2n}\bigl(\{ p_i \}_{i=1}^{2n}\bigl)$ represents the proper $2n$-point vertex of composite fermions, 
defined by 
\beq
\Gamma_{2n}\bigl(\{ p_i \}_{i=1}^{2n}\bigl)\!\!\!\!&=&\!\!\!\!
-\frac{1}{N^{n-1}}
\int dq \prod_{j=1}^{n}  D\biggl(\:\sum_{k=1}^{j-1}p_{2k-1}-\sum_{k=1}^{j-1}p_{2k}+q\biggl)\: 
S\biggl(\:\sum_{k=1}^{j}p_{2k} -\sum_{k=1}^{j-1}p_{2k-1}-q\biggl) 
 \: ,
\eeq  
with a set of momenta $\{p_i\}_{i=1}^{2n}=\{p_1, p_2 , \dots , p_{2n}\}$.   

\newpage

\section{Derivative expansion of the inverse propagator for composite fermions at $T=0$}
\label{Cal_G}
 
We will give details on the derivation of Eq.(\ref{low_G}) and Eqs.(\ref{coeff_a})-(\ref{coeff_d}) 
in the derivative expansion of $G^{-1}(p)$ at zero temperature. 
We can formally perform a real-time analysis by replacing our Matsubara frequency $i\omega$ 
with a continuous energy variable $E$ at zero temperature. 
Using a four-momentum in real-time formalism; $({\bf p}, E)$, 
we have a real-time form of Eq.(\ref{F_prop2}) at zero temperature, 
\beq
G^{-1}({\bf p}, E)&=&
\frac{\mR}{2\pi \abf} - \int\!\frac{d {\bf q}}{(2\pi)^3}\left\{ \frac{1}{ \varepsilon ( {\bf q} )}-
 \frac{1}{\xif({\bf p}-{\bf q})+\xib({\bf q})-E}
 \right\} \nonumber\\
&=& \frac{\mR}{2\pi \abf} - \int\!\frac{d {\bf q}}{(2\pi)^3}\left\{ \frac{1}{ \varepsilon ( {\bf q} )}-
 \frac{1}{\varepsilon ( {\bf q} )- {\bf p}\cdot{\bf q}/\mf+ {\bf p}^2/(2\mf)- E-(\mub+\muf)}
 \right\} \nonumber\\
&=&  \frac{\mR}{2\pi \abf} - I_1\: .
\label{real_G}
\eeq
Here  we used the fact that $n_b({\bf q})$ and $n_f({\bf p}-{\bf q})$  vanishes at $T=0$ 
under our assumption: $\mub<0$ and $\muf<0$, and 
we denoted an integral in Eq.(\ref{real_G}) by $I_1$, which becomes 
\beq
I_1&=&\int\!\frac{d {\bf q}}{(2\pi)^3}\left\{ \frac{1}{ \varepsilon ( {\bf q} )}-
 \frac{1}{\varepsilon ( {\bf q} )- {\bf p}\cdot{\bf q}/\mf+p^2/(2\mf)- E+|\mu|}
 \right\} \nonumber\\
&=& \frac{1}{4\pi^2}\int_0^{\Lambda}q^2dq\int_{-1}^1d\cos\theta
\left\{ \frac{1}{ \varepsilon ( {\bf q} )}-
 \frac{1}{\varepsilon ( {\bf q} )- pq\cos\theta/\mf+A}
 \right\}\nonumber\\
&=&  \frac{\mR\Lambda}{\pi^2}- \frac{1}{4\pi^2}\int_0^{\Lambda}q^2dq\:
\frac{\mf}{pq}\ln \biggl|\frac{pq/\mf +\varepsilon ( {\bf q} )+A }{-pq/\mf +\varepsilon ( {\bf q} )+A} \biggl| \:, 
\label{def_I1}
\eeq
with $p=|{\bf p}|$, $q=|{\bf q}|$, $\mu=\mub+\muf(<0)$, and $A= p^2/(2\mf)- E+|\mu|$. 
As we will see below, $I_1$ yields a finite value even in the limit of $\Lambda\to\infty$. 
We can rewrite terms in the logarithm in Eq.(\ref{def_I1}) as 
\beq
\frac{pq}{\mf} +\varepsilon ( {\bf q} )+A 
&=&\frac{1}{2\mR}\biggl( q^2 +2 \frac{\mR}{\mf}pq \biggl)+\frac{p^2}{2\mf}-E +|\mu|
\nonumber\\
&=&\frac{1}{2\mR}\bigl\{(q+\alpha p)^2 +2\mR A'\bigl\} \: ,
\eeq
and 
\beq
-\frac{pq}{\mf} +\varepsilon ( {\bf q} )+A 
&=&\frac{1}{2\mR}\bigl\{(q-\alpha p)^2 +2\mR A'\bigl\}\: , 
\eeq
with a mass-ratio parameter $\alpha=\mR/\mf$, and $ A'=p^2/[2(\mb+\mf)]-E+|\mu|$. 
Then, Eq.(\ref{def_I1}) reads 
\beq
I_1&=&
\frac{\mR\Lambda}{\pi^2}- \frac{\mf}{4\pi^2 p}\int_0^{\Lambda}dq\: q
\ln \biggl|\frac{(q+\alpha p)^2 +2\mR A'}{(q-\alpha p)^2 +2\mR A'} \biggl| \: .
\eeq
By employing the following integral formula: 
\beq
\lefteqn{ \int dx \: x\ln \biggl| \frac{(x+a)^2+b}{(x-a)^2+b} \biggl|} \\
&=&\biggl(\frac{x^2-a^2+b}{2}\biggl)\ln \biggl| \frac{(x+a)^2+b}{(x-a)^2+b} \biggl|
+\: 2 a x -2a\sqrt{|b|} \biggl\{\tan^{-1}\biggl(\frac{x+a}{\sqrt{|b|}}\biggl)+ \tan^{-1}\biggl(\frac{x-a}{\sqrt{|b|}}\biggl)\biggl\}
\: ,
\nonumber
\eeq
we can perform the integration in Eq.(\ref{def_I1}), 
\beq
I_1&=&
\frac{\mR\Lambda}{\pi^2}
-\frac{\mf}{4\pi^2 p}  \biggl\{\frac{\Lambda^2-(\alpha p)^2+2\mR A'}{2}\biggl\}
\ln \biggl| \frac{(\Lambda+\alpha p)^2+2\mR A'}{(\Lambda-\alpha p)^2+2\mR A'} \biggl|
\nonumber\\
&&
-\frac{\mf}{4\pi^2 p}  2 \alpha p \Lambda 
+\frac{\mf}{4\pi^2 p}  2\alpha p\sqrt{|2\mR A'|}
\biggl\{ \tan^{-1}\biggl(
\frac{\Lambda+\alpha p}{\sqrt{|2\mR A'|}}\biggl)+ \tan^{-1}\biggl(\frac{\Lambda-\alpha p}{\sqrt{|2\mR A'|}}
\biggl)\biggl\}
\nonumber\\
&=&\frac{\mR\Lambda}{\pi^2}
-\frac{\mf}{4\pi^2 p}  \biggl(\frac{\Lambda^2-\alpha^2 p^2+2\mR A'}{2}\biggl)
\ln \biggl| \frac{(\Lambda+\alpha p)^2+2\mR A'}{(\Lambda-\alpha p)^2+2\mR A'} \biggl|
\nonumber\\
&&
-\frac{\mR  \Lambda }{2\pi^2 }
+\frac{\mR}{2\pi^2 } \sqrt{|2\mR A'|}
\biggl\{ \tan^{-1}\biggl(
\frac{\Lambda+\alpha p}{\sqrt{|2\mR A'|}}\biggl)+ \tan^{-1}\biggl(\frac{\Lambda-\alpha p}{\sqrt{|2\mR A'|}}
\biggl)\biggl\}\: .
\label{I_1_2}
\eeq

Noting that the cutoff $\Lambda$ is the largest scale in our model 
and also that we are interested in the low-energy and low-momentum regime, 
let us expand the logarithm and arctangent functions in Eq.(\ref{I_1_2}). 
As for the logarithm function, 
the power-series formula, $\ln(1+x)=\sum_{n=1}^{\infty}(-1)^{n-1}x^n/n$, gives 
\beq
\lefteqn{  \biggl(\frac{\Lambda^2-\alpha^2 p^2+2\mR A'}{2}\biggl)
\ln \biggl| \frac{(\Lambda+\alpha p)^2+2\mR A'}{(\Lambda-\alpha p)^2+2\mR A'} \biggl| } \nonumber\\
&=&\frac{\Lambda^2}{2} \biggl(1-\frac{\alpha^2 p^2-2\mR A}{\Lambda^2}'\biggl)
\ln \biggl| \frac{1+2\alpha p/\Lambda+(\alpha^2p^2+2\mR A')/\Lambda^2}
{1-2\alpha p/\Lambda+(\alpha^2p^2+2\mR A')/\Lambda^2} \biggl|
\nonumber\\
&=&\frac{\Lambda}{2} \biggl(1-\frac{\alpha^2 p^2-2\mR A}{\Lambda^2}'\biggl)
\biggl\{4\alpha p +O\biggl(\frac{\alpha p(\alpha^2p^2+2\mR A')}{\Lambda^2}\biggl) \biggl\}
\nonumber\\
&=&2\alpha p \Lambda \:+\: O\biggl(\frac{\alpha p(\alpha^2p^2\pm2\mR A')}{\Lambda}\biggl) \:, 
\eeq
while the formula for the arctangent function: $\tan^{-1}(x)=\pi/2-\tan^{-1}(1/x)$, yields 
\beq
\tan^{-1}\biggl(
\frac{\Lambda+\alpha p}{\sqrt{|2\mR A'|}}\biggl)\:+\: \tan^{-1}\biggl(\frac{\Lambda-\alpha p}{\sqrt{|2\mR A'|}}
\biggl)
&=&\pi \:+\:O\biggl(\frac{\sqrt{|2\mR A'|}}{\Lambda\pm\alpha p}\biggl) \:,  
\eeq
which finally give an explicit form of $I_1$ within the derivative expansion, 
\beq
I_1&=&
\frac{\mR\Lambda}{\pi^2}
-\frac{\mR  \Lambda }{2\pi^2 }
-\frac{\mR  \Lambda }{2\pi^2 }
+\frac{\mR}{2\pi } \sqrt{|2\mR A'|}
\:+\: O\biggl(\frac{\alpha p(\alpha^2p^2\pm2\mR A')}{\Lambda}\biggl)
\:+\:O\biggl(\frac{\sqrt{|2\mR A'|}}{\Lambda\pm\alpha p}\biggl) \nonumber\\
&=&\frac{(2\mR)^{3/2} }{4\pi } \sqrt{|A'|}
\:+\: O\biggl(\frac{\alpha p(\alpha^2p^2\pm2\mR A')}{\Lambda}\biggl)
\:+\:O\biggl(\frac{\sqrt{|2\mR A'|}}{\Lambda\pm\alpha p}\biggl)\: .
\label{der_I_1}
\eeq
Substituting Eq.(\ref{der_I_1}) into Eq.(\ref{real_G}) yields
\beq
G^{-1}({\bf p}, E)&\simeq&
\frac{\mR}{2\pi \abf} - \frac{(2\mR)^{3/2} }{4\pi } \sqrt{\:|\mu|+\frac{{\bf p}^2}{2(\mb+\mf)} -E\:}\: ,
\eeq
which finally gives 
the derivative expansion of $G^{-1}$ in terms of $\{{\bf p}^2/[2(\mb+\mf)] -E\}/|\mu|$ 
under the assumption sated in the main text, 
\beq
G^{-1}({\bf p}, E)\!\!&\simeq&\!\!
\frac{\mR}{2\pi \abf} - \frac{(2\mR)^{3/2} }{4\pi }
\sqrt{|\mu|}\biggl(1+\frac{{\bf p}^2/[2(\mb+\mf)]-E}{|\mu|}\biggl)^{1/2}
\nonumber\\
\!\!&\simeq&\!\!
\frac{\mR}{2\pi \abf}\bigl(1-\abf\sqrt{2\mR |\mu|}\;\bigl)
\:-\: \frac{1}{2\pi}\frac{\mR}{2(\mb+\mf)}\sqrt{\frac{\mR}{2|\mu|}}\;{\bf p}^2\: 
\:+\: \frac{\mR}{2\pi}\sqrt{\frac{\mR}{2|\mu|}}\; E\: . 
\eeq
The above expression corresponds to 
Eq.(\ref{deriv_G}) with the coefficients $a,c,d$ given by Eq.(\ref{coeff_a})-(\ref{coeff_d}). 





\begin{thebibliography}{99}

\bibitem{BBP67} J.~Bardeen, G.~Baym, and D.~Pines, 
Phys. Rev. {\bf 156}, 207 (1967). 

\bibitem{BHS00} M.~J.~Bijlsma, B.~A.~Heringa, and H.~T.~C.~Stoof, 
Phys. Rev. A {\bf 61}, 053601 (2000). 

\bibitem{HPSV00} H.~Heiselberg, C.~J.~Pethick, H. Smith, and L.~Viverit, 
Phys. Rev. Lett. {\bf 85}, 2418 (2000).  

\bibitem{SSSYD05} A.~Storozhenko, P.~Schuck, T.~Suzuki, H.~Yabu, and J.~Dukelsky, 
Phys. Rev. A {\bf 71}, 063617 (2005). 


\bibitem{KBEK04} M.~Yu.~Kagan, I.~V.~Brodsky, D.~V.~Efremov, and A.~V.~Klaptsov, 
Phys. Rev. A {\bf 70}, 023607 (2004).

\bibitem{BHTY08} G.~Baym, T.~Hatsuda, M.~Tachibana, and N.~Yamamoto, 
 J. Phys. G {\bf 35}, 104021 (2008). 

\bibitem{MBH09} K.~Maeda, G.~Baym, and T.~Hatsuda, 
Phys. Rev. Lett. {\bf 103}, 085301 (2009). 

\bibitem{HM09} T.~Hatsuda and K.~Maeda, 
arXiv:0912.1437[hep-ph]. 

Chapter of the book: {\it Understanding Quantum Phase transitions}, edited by L.~D.~Carr 
(CRC Press, Taylor and Francis, 2010). 

\bibitem{FES58} H.~Feshbach, Ann. Phys. (N.Y.) {\bf 5}, 357 (1958); {\bf 19}, 287 (1962).  

\bibitem{KGJ06} T.~K\"{o}hler, K.~G\'{o}ral, and P.~S.~Julienne, 
Rev. Mod. Phys. {\bf 78}, 1311 (2006). 

\bibitem{BDZ08} I.~Bloch, J.~Dalibard, and W.~Zwerger, Rev. Mod. Phys. {\bf 80}, 885 (2008).  

\bibitem{OSP06} C.~Ospelkaus, S.~Ospelkaus, L.~Humbert, P.~Ernst, K.~Sengstock, and K.~Bongs,  
Phys. Rev. Lett. {\bf 97}, 120402 (2006).

\bibitem{ZIR08} J.~J.~Zirbel, K.~-K.~Ni, S.~Ospelkaus, J.~P.~D'Incao, C.~E.~Wieman, J.~Ye, and D.~S.~Jin, 
Phys. Rev. Lett. {\bf 100}, 143201 (2008). 

\bibitem{BBZ00} G.~Baym, J.-P.~Blaizot, and J.~Zinn-Justin, 
Europhys. Lett. {\bf 49}, 150 (2000).

\bibitem{AMT01} P.~Arnold, G.~Moore, and B.~Tom\'{a}\v{s}ik,
Phys. Rev. A {\bf 65}, 013606 (2001). 

\bibitem{NS07} P.~Nikoli\'{c} and S.~Sachdev, 
Phys. Rev. A {\bf 75}, 033608 (2007).

\bibitem{VSR07} M.~Y.~Veillette, D.~E.~Sheehy, and L.~Radzihovsky, 
Phys. Rev. A {\bf 75}, 043614 (2007).

\bibitem{AB08} H.~Abuki and T.~Brauner, 
Phys. Rev. A {\bf 78}, 125010 (2008).

\bibitem{MZJ2003} M.~Moshe and J.~Zinn-Justin, 
Phys. Rept. {\bf 385}, 69 (2003). 


\bibitem{PS2008} C. J. Pethick and H. Smith, \textit{Bose-Einstein Condensation in Dilute Gases} 
(Cambridge University Press, Cambridge, 2008). 


\bibitem{RHET02} L.~J.~Abu-Raddad, A.~Hosaka, D.~Ebert, and H.~Toki, 
Phys. Rev. D {\bf 66}, 025206 (2002). 

\bibitem{NO1988}  J.~W.~Negele and H.~Orland, 
{\it Quantum Many-particle systems} 
(Westview Press, Boulder, 1988). 

\bibitem{SRE93} C.~A.~R.~S\'{a} de Melo, M.~Randeria, and J.~R.~Engelbrecht, 
Phys. Rev. Lett. {\bf 71}, 3202 (1993).

\bibitem{HAU93} R.~Haussmann, 
Z. Phys. B: Condens. Matter {\bf 91}, 291 (1993).

\bibitem{PS96} F.~ Pistolesi and G.~C.~Strinati, 
Phys. Rev. B {\bf 53}, 15168 (1996). 

\bibitem{PS00} P.~ Pieri and G.~C.~Strinati, 
Phys. Rev. B {\bf 61}, 15370 (2000). 

\bibitem{BCS57} J.~Bardeen, L.~N.~Cooper, and J.~R.~Schrieffer, 
Phys. Rev. {\bf 106}, 162 (1957).

\bibitem{GMB61} L.~P.~Gor'kov and T.~K.~Melik-Barkhudarov, 
Sov. Phys. JETP {\bf 13}, 1081 (1961).

\bibitem{PSS04} D.~S.~Petrov, C.~Salomon, and G.~V.~Shlyapnikov, 
Phys. Rev. Lett. {\bf 93}, 090404 (2004); 
Phys. Rev. {\bf A71}, 012708 (2005); 
J. Phys. B, {\bf 38}, S645 (2005); 
arXiv:0810.1949 [cond-mat]. 

\bibitem{BKKCL06} I.~V.~Brodsky, M.~Yu.~Kagan, A.~V.~Klaptsov, R.~Combescot, and X.~Leyronas, 
Phys. Rev. {\bf A73}, 032724 (2006). 

\bibitem{LG06}
J.~Levinsen and V.~Gurarie, 
Phys. Rev. {\bf A73}, 053607 (2006).

\bibitem{DHJ03} D.~J.~Dean and M.~Hjorth-Jensen, 
Rev. Mod. Phys. {\bf 75}, 607 (2003). 

\bibitem{ASRS08} M. G. Alford, A. Schmitt, K. Rajagopal, and T. Schafer, 
Rev. Mod. Phys. {\bf 80}, 1455 (2008). 


\end{thebibliography}
\end{document}